\documentclass{raa}           
\usepackage{graphicx,times}
\usepackage{natbib}
\usepackage{amssymb,amsmath}
\bibpunct{(}{)}{;}{a}{}{,}

\begin{document}

\title{The temperature of IGM at high redshifts: shock heating and high mach problem}

 \volnopage{ {\bf 20XX} Vol.\ {\bf X} No. {\bf XX}, 000--000}
   \setcounter{page}{1}

   \author{Junyi Jia\inst{1,2}, Weishan Zhu\inst{3}, Liang Gao\inst{1,2,4}, Long-Long Feng
      \inst{3,5}
   }

   \institute{ Key Laboratory for Computational Astrophysics, National Astronomical Observatories, Chinese Academy of Sciences, Beijing, 100012, China
China; {\it jyjia@nao.cas.cn}\\
     \and
      School of Astronomy and Space Science, University of Chinese Academy of Sciences, Beijing 10039, China\\
	\and
			School of Physics and Astronomy,  Sun Yat-Sen University, Zhuhai 519082, China\\
\and 
	Institute of Computational Cosmology, Department of Physics, University of Durham, South Road, Durham DH1 3LE, UK\\
    \and
    Purple Mountain Observatory, CAS, Nanjing, 210008, China\\
\vs \no
   {\small Received 20XX Month Day; accepted 20XX Month Day}
}

\abstract{The thermal history of cosmic gas in the Dark Ages remains largely unknown. It is important to quantify the impact of relevant physics on the IGM temperature between $z=10$ and $z \sim 30$, in order to interpret recent and oncoming observations, including results reported by EDGES. We revisit the gas heating due to structure formation shocks in this era, using a set of fixed grid cosmological hydrodynamical simulations performed by three different codes. In all our simulations, the cosmic gas is predicted to be in multiphase state since $z>30$. The gas surrounding high density peaks gradually develops a relation more sharp than $T \propto \rho^{2/3}$, approximately $T \propto \rho^{2}$, from $z=30$ to $z=11$, might due to shock heating. Meanwhile, the gas in void region tends to have a large local mach number, and their thermal state varies significantly from code to code. In the redshift range $11-20$, the mass fraction of gas  shock heated above the CMB temperature in our simulations is larger than  previous semi-analytical results by a factor of 2 to 8. At $z=15$, the fraction varies from $\sim 19\%$ to $52 \%$ among different codes. Between $z=11$ and $z=20$, the gas temperature $<1/T_{\rm{K}}>_M^{-1}$ is predicted to be $\sim 10-20$ K by two codes, much higher than the adiabatic cooling model and some previous works. However, in our simulations performed by RAMSES, $<1/T_{\rm{K}}>_M^{-1}$ is predicted to be even below the temperature required to explain result of the EDGES. Given the fact that different codes give different predictions, currently, it seems a challenge to make solid prediction on the temperature of gas at $z \sim 17$ in simulations.
}

   \authorrunning{Junyi Jia, et al.}            
   \titlerunning{IGM at high z: shock heating}  
   \maketitle

%
\section{Introduction}
In the  $\Lambda \rm{CDM}$ cosmology, the baryonic matter accounts for around $5 \%$ of the mass budget of the universe, based on observations including the cosmic microwave background, supernovae, and galaxy clustering(e.g., \citealt{Planck2014}). Despite the small mass fraction that baryons contribute, their states are crucial to our understanding of the universe as almost all of the observed light comes from baryons. The properties of the baryons since the complete of re-ionization have been comprehensively investigated by many theoretical and observational works. The Lyman-$\alpha$ lines in the spectrum of Quasars have revealed most of the intergalactic medium(IGM) and their properties in the range of $1<z<5.5$(\citealt{Gunn1965}; \citealt{Cen1994}; \citealt{Hu1995}; \citealt{Rauch1997}). Although as much as $\sim 30-40 \%$  of the baryons have  not been observed at $z<1$(\citealt{Shull2012}), numerical simulations predict that those missing baryons should be in the state of Warm and Hot IGM(WHIM), residing in filamentary structures(\citealt{Cen1999}; \citealt{Dave2001};\citealt{Cui2019}). Many observational methods have been proposed and some experiments are being conducted to search the WHIM(\citealt{Bregman2007}; \citealt{McQuinn2016}; \citealt{Nicastro2018}). 
 
On the other hand, the properties of IGM during re-ionization and at even earlier times, i.e., in the redshift range of $10 \lesssim z \lesssim 100-200$, remain largely unknown. So far,  a general sketch of the thermal evolution of the IGM in this era has been outlined by some theoretical studies. After the thermal decoupling from the CMB photons at $z\sim 200$, the baryonic gas is anticipated to firstly cool adiabatically as $T \propto (1+z)^{-2}$. Later on, the collapse of structures would reheat the cosmic gas. Shocks generated by structure formation can effectively transform the kinetic energy to internal energy and further contribute to the reheating of gas(\citealt{Sunyaev1972}; \citealt{Furlanetto2004}; \citealt{Gnedin2004}; \citealt{Furlanetto2006}; \citealt{McQuinn2012}). Also, the emergence of the first generation of stars and galaxies would gradually produce radiation background that lead to the re-ionization of the IGM at $z<10-30$. In addition, it has been proposed that in the early universe the cosmic gas moves supersonically with respect to the dark matter, and the initial velocity differences would also trigger shocks and reheat the baryons(e.g., \citealt{Tseliakhovich2010}; \citealt{OLeary2012}). 

Redshifted 21cm observations have long been proposed to be a probe of the properties of the cosmic gas in the pre-reionization era (e.g., \citealt{Madau1997}). Very recently, \citet{Bowman2018} report their result from the Experiment to Detect the Global Epoch of  Reionization Signature (EDGES), claiming the detection of a strong 21-cm absorption signal with respect to the CMB at $z \sim17$. \citet{Bowman2018} concludes that in order to explain the EDGS features, the temperature of gas, $T_g$, should be less than 3.2K if the spin temperature of hydrogen is fully coupled to $T_g$ and the radiation temperature is solely determined by CMB. Such a gas temperature is colder than the expected value in adiabatic cooling scenario, i.e., 9.3 K at $z = 20$, and 5.4 K at $z = 15$. Alternatively, the temperature of background radiation temperature should be hotter than expected. Interactions between baryons and dark matter(\citealt{Barkana2018}), as well as several other mechanisms have been proposed to interpret the EDGES signal(e.g. \citealt{Feng2018}). 

For an accurate model that can explain the strong absorption reported by EDGES, a precise knowledge of the gas temperature $T_g$ is needed. The spin temperature of hydrogen is coupled to the gas temperature via atomic collisions or the Wouthuysen-Field effect. However, as  mentioned above, the IGM temperature at $z \sim 20$ is subject to multi-physical processes. Currently, there are many uncertainties in those processes. For instance, the formation and evolution of the first generations of stars have significant impact on reheating the cosmic gas  at $z < \sim 10$, while they are poorly constrained. Even at high redshifts $z >\sim 10$ when the Lyman-alpha and X-ray heating by stars may be weak, there exists a notable difference on predicting the impact of structure formation shocks on the heating of baryonic gas in previous works. Order of unity differences can be found in the literature. 

Using cosmological hydrodynamical simulations, \citet{Gnedin2004} demonstrated that gas heating at $z\sim 10-17$ was mainly due to structure formation shocks. Based on an analytic model, \citealt{Furlanetto2004} predicted that around $10\%$ of the gas is heated by large-scale structure shocks in the redshift range $10<z<20$, and the mass weighted mean temperature of shock heated gas increase from $\sim 200$K at $z=20$ to $\sim 5000$K at $z=10$. Using a semi-analytical model, \citet{Furlanetto2006} further estimated that about $0.1 \%, 3\%, 25\%$ of the gas would be shock heated to above the temperature of CMB at redshift $30, 20, 10$ respectively. \citet{McQuinn2012} used two different hydro-dynamical codes, GADGET and Enzo, to quantify the impact of shock heating on the thermal history of the cosmic gas. They found that the gas temperature in their simulations deviate from the adiabatic model by $\approx 5\%-20\%$ in the redshift range $10-20$. However, in disagreement with \citet{Gnedin2004}, they found that gas heating due to structure formation shocks is not the dominant, the Compton heating is equally important. In addition to the discrepancies discussed above, the properties of the structure formation shocks at high redshifts, such as the spacial distribution, statistics of mach number and frequency, have been less tackled in the literature. 

Considering the significant discrepancies discussed above, in this paper, we revisit the heating of the IGM by structure formation shocks at high redshifts with cosmological hydrodynamical simulations. The organization of the paper is as follows. We introduce our simulation methods and details in section 2. The thermal state of IGM and the properties of schocks in the redshift range of $11<z<30$ in our simulations are presented in section 3. Finally, we summarize our findings and discuss our results in section 4.

\section{Codes and Simulations}

\subsection{Codes}
During the dark ages, gas temperature is quite low due to the expansion of the  universe. As a result, when gas moves from under-dense to over-dense regions, it is easy and common to become supersonic. It has been a long standing problem to resolve such gas dynamics accurately in cosmological hydrodynamical simulations, which is called as the high mach problem. More specifically, as the internal thermal energy is much smaller than the kinetic energy for supersonic flow, small numerical errors on the total energy and kinetic energy can lead to the thermal energy, and the gas temperature change significantly. Therefore, high precision hydrodynamical solvers are needed to trace gas motion. In this study we choose two types of numerical schemes to solve  hydrodynamical equations to see how results are sensitive to different numerical schemes. One is the Pieceweise Linear Method(PLM), which is a second-order Godunov method(\citealt{Toro1997}). For this scheme, we adopt a popular grid-based cosmological hydrodynamical simulation code RAMSES(\citealt{Teyssier2002}). RAMSES uses a "one-way interface" scheme to solve the Poisson equation to obtain the gravitational potential at grids.  More specifically, the Poisson equation is solved with the Gauss-Seidel (GS) relaxation method. For more details of the code, we refer the reader to \citet{Teyssier2002}. The other scheme used in this study is the fifth order weighted essentially non-oscillatory (WENO) scheme firstly developed by \citet{Jiang1996}. The WENO scheme has been implemented in a Cosmological simulation code WIGEON(\citealt{Feng2004}), which uses standard particle mesh(PM) method to solve the gravitational potential. 

Note, in addition to the hydrodynamical solver, there exists other differences between RAMSES and WIGEON. For instance, the Poisson equation solver, and the treatment of source terms in the right hand of hydrodynamical equations are not completely consistent with each other; both codes also  adopt different temporal discretization schemes. Therefore, any difference between simulations run with these two codes will be result of the combined factors. In order to make a more direct comparison of hydrodynamical schemes, we introduce the third code which solves the hydrodynamics with PLM method and uses second-order midpoint time integration as in RAMSES, while the gravity solver and the treatment of sources terms related to gravity in hydrodynamical equations are identical to WIGEON. We denote this code as 'PLM-PM' in the following context. 
 
 To prevent the gas temperature and density artificially fall below zero in numeric, we adopt the positive preserving WENO scheme in WIGEON. The scheme was firstly introduced by \citet{Zhang2012}, and its performance on cosmological simulations was discussed in \citet{Zhu2013}. For RAMSES and PLM-PM, a method similar to \citet{Bryan1995} is used to avoid negative energy. Namely, both the total and the internal energy of gas at each grid cell is tracked and updated at each time step during the entire simulation. In addition, floors of gas density and temperature, $\rho_{g, floor}=10^{-12}, T_{floor}=10^{-12}$, are enforced in our simulations run with RAMSES and PLM-PM.

\subsection{Simulations}

\begin{table*}
	\centering
	\caption{The simulations performed in this work. The side length of cubic box is $25 h^{-1}$ Mpc for all the simulations.}
	\label{tab:simu_table}
	\begin{tabular}{|c|c|c|c|c|} 
		\hline
		simulation & hydro solver & heating, cooling & grids/particles & dark matter particle mass\\ 
		\hline

	    WIGEON-ada-512 & WENO & adiabatic & $512^3$ & $1.0 \times 10^{7}M_{\odot}$\\
		PLM-PM-ada-512 & PLM & adiabatic &$512^3$ & $1.0 \times 10^{7}M_{\odot}$\\
		RAMSES-ada-512 & PLM  & adiabatic &$512^3$ & $1.0 \times 10^{7}M_{\odot}$\\
		WIGEON-uvc-256 & WENO & Compton, UV, cooling & $256^3$ & $8.2 \times 10^{7}M_{\odot}$\\
		PLM-PM-uvc-256 & PLM & Compton, UV, cooling &$256^3$ & $8.2 \times 10^{7}M_{\odot}$\\
		RAMSES-uvc-256 & PLM  & Compton, UV, cooling &$256^3$ & $8.2 \times 10^{7}M_{\odot}$\\
		WIGEON-uvc-512 & WENO & Compton, UV, cooling & $512^3$ & $1.0 \times 10^{7}M_{\odot}$\\
		PLM-PM-uvc-512 & PLM & Compton, UV, cooling &$512^3$ & $1.0 \times 10^{7}M_{\odot}$\\
		RAMSES-uvc-512 & PLM  & Compton, UV, cooling &$512^3$ & $1.0 \times 10^{7}M_{\odot}$\\
		WIGEON-uvc-1024 & WENO & Compton, UV, cooling & $1024^3$ &$1.3 \times 10^{6}M_{\odot}$\\
		PLM-PM-uvc-1024 & PLM & Compton, UV, cooling &$1024^3$ & $1.3 \times 10^{6}M_{\odot}$\\
		RAMSES-uvc-1024 & PLM  & Compton, UV, cooling &$ 1024^3$ & $1.3 \times 10^{6}M_{\odot}$\\
		\hline
	\end{tabular}
\end{table*}

We perform two sets of cosmological hydrodynamical simulations with three codes introduced in the last subsection. One set is non-radiative, and the other set includes radiative processes, namely, the Compton heating and cooling by CMB photons, and a uniform UV background extrapolated from \citealt{Haardt1996}, and radiative cooling. For the latter set, the UV background is switched on at $z=15$, and the radiative processes are modeled by following \citet{Theuns1998} assuming a pristine gas composition($X=0.76, Y=0.24$). Note, physical parameters related to the Compton process, UV  heating and radiative cooling are set to be the same in three different codes.

We evolve all our simulations in a periodic cubic box with a side length $25 h^{-1}$Mpc. Comparing to two similar studies in terms of the simulation volume,  ours is larger than \citet{Gnedin2004} but is smaller than \citet{McQuinn2012}. For the non-radiative runs, we evolve simulations with a number of $512^3$ grids and dark matter particles respectively. For the radiative cooling simulations, we run with three different resolutions to probe the effect of resolution. One with the same resolution as the non-radiative run and the other with 8 times higher/lower resolution, i.e., $1024^3$/$256^3$ grids and particles. The spacial and mass resolution are $24.4 h^{-1}$kpc and $1.30 \times 10^{6}M_{\odot}$ respectively for our highest resolution runs. The physics and parameters of our simulations are listed in Table~\ref{tab:simu_table}. All simulations adopt the Planck cosmology, i.e., $\Omega_{m}=0.317, \Omega_{\Lambda}=0.683,h=0.671,\sigma_{8}=0.834, \Omega_{b}=0.049$, and $n_{s}=0.962$(\citealt{Planck2014}).  All our simulations are started from $z=99$ and evolved to $z=0$ .  The initial gas temperature of all simulations is set to be the same, i.e. $T_g(z=99) = 152$K.

\begin{figure}
\begin{center}
\includegraphics[width=0.6\columnwidth]{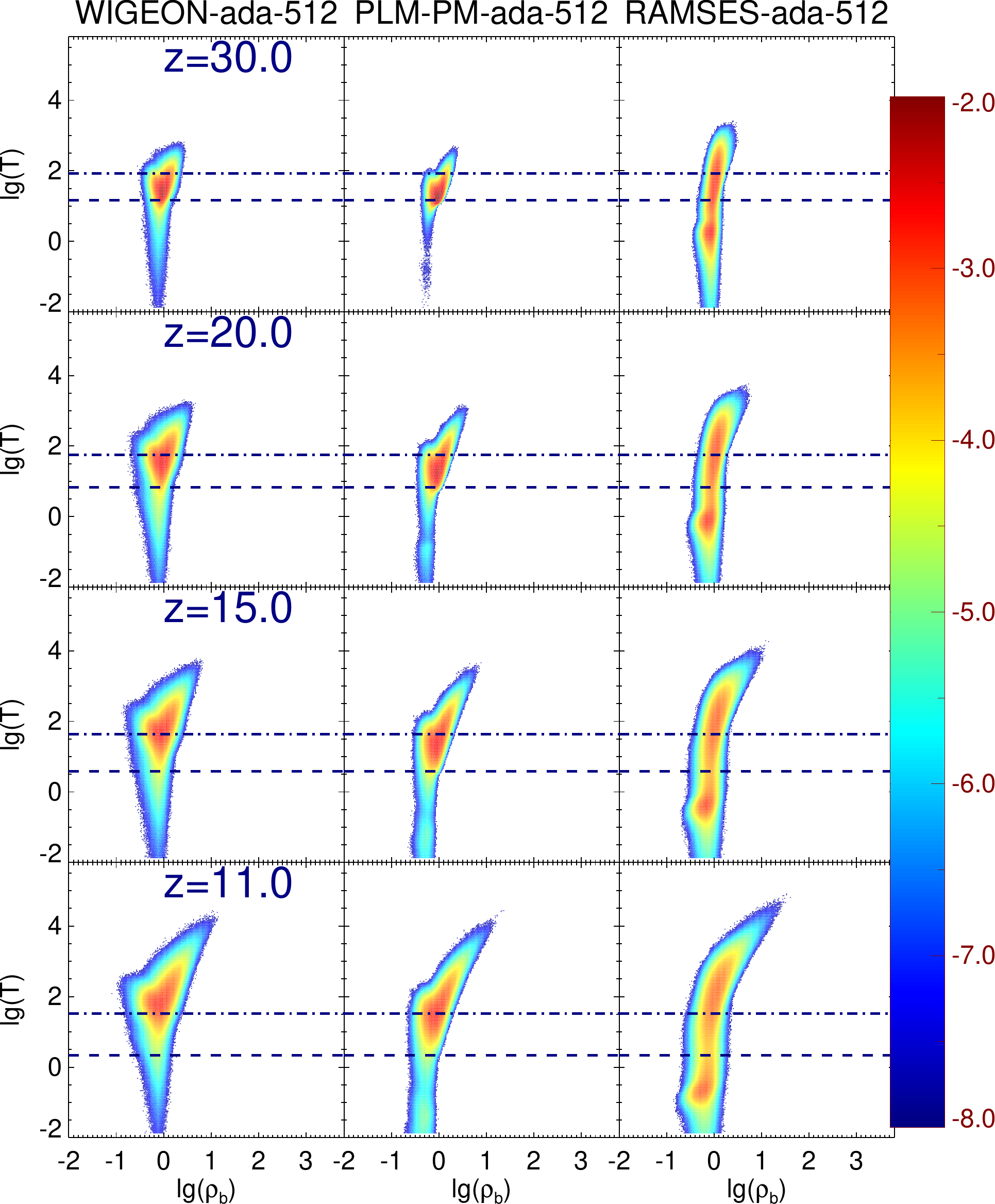}
\includegraphics[width=0.6\columnwidth]{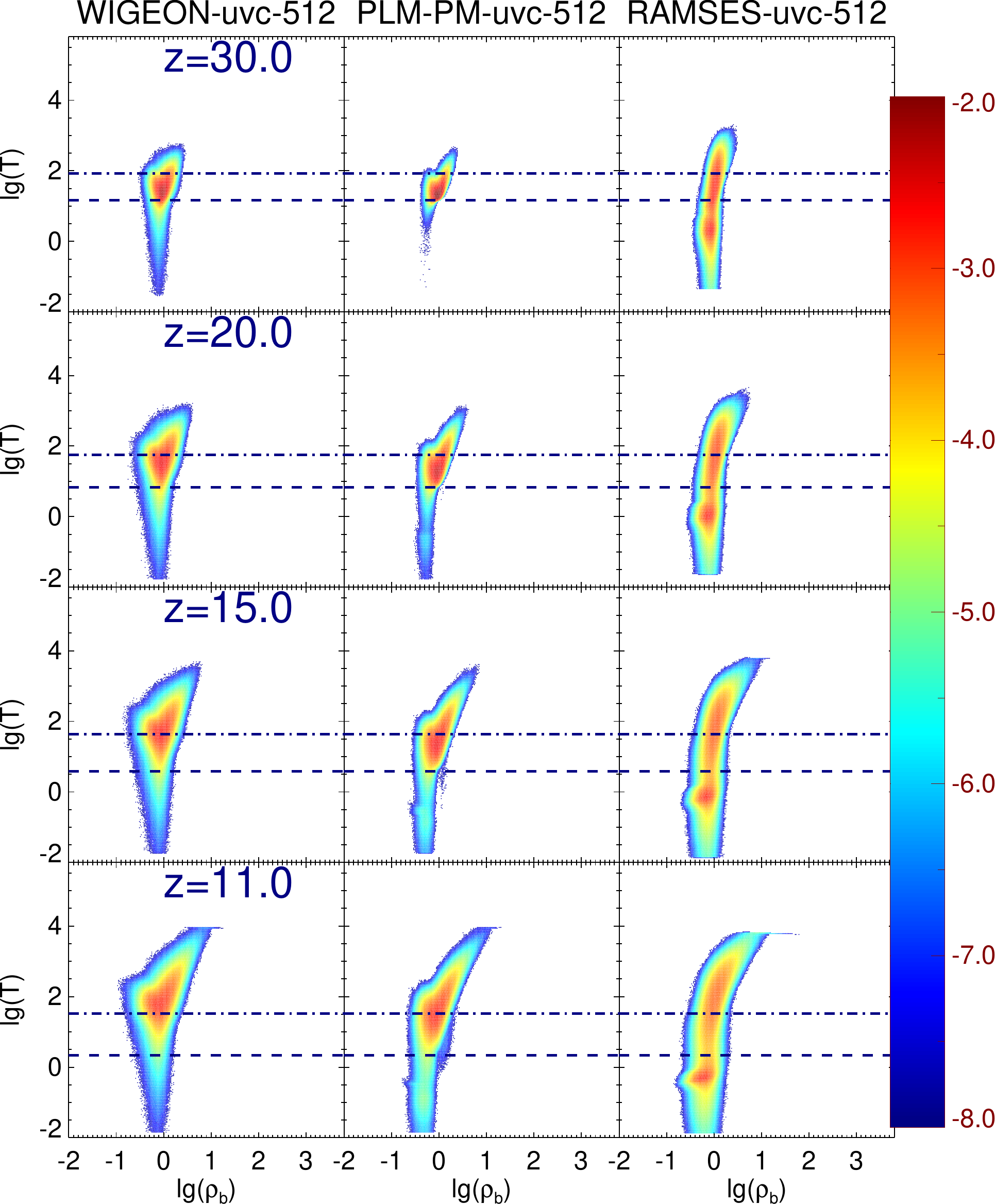}
\caption{Top: The volume-weighted distribution of the cosmic gas in the density-temperature plane at redshift $30, 20, 15, 11$ in our non-radiative simulations; Left, middle and right column indicate simulations run with WIGEON, PLM-PM, and RAMSES, respectively; Dotted-dashed lines in each plot indicate the CMB temperature at corresponding redshifts; Dashed lines indicate the expected gas temperature due to expansion. Bottom: same as the top panel, but for simulations including Compton process, a uniform UV background, and radiative cooling.}
\label{fig:vol_dis_ada}
\end{center}
\end{figure}

\section{Results}

\subsection{Evolution of gas phases in $(\rho, T)$ space}

We start our analysis from the thermal history of the gas in the non-radiative simulations. In these simulations, the gas is only heated by gravitational collapse of structure formation. In the top panel of Fig.~\ref{fig:vol_dis_ada} we show the volume-weighted distribution of cosmic gas in the density-temperature phase plane at redshift $z=30, 20, 15, 11$. Apparently, the gas in all the three simulations starts to develop to a multi-phase state at $z>30$, because of the combined effects of heating by structure formation shocks and cooling by cosmic expansion. The evolution of the density-temperature plane predicted by three codes show evident differences. The distribution of gas in the simulations run with WIGEON and RAMSES is moderately more extended in $(\rho, T)$ space than that in the one run with PLM-PM along.

The different gravity and hydrodynamical solvers should have led to different paces of structure collapse, and different intensities of shock heating in different codes. We show the density power of both dark matter and gas in the Appendix A. Notable discrepancies can be observed at scales $\sim 1$ Mpc at $z=11$. Giving the high mach properties of gas at this era, the accuracy on gas motions and energy of different numerical solvers would lead to remarkable numerical uncertainty and differences in gas temperature between different codes. For instance, the treatment of the gravitational sources terms is crucial to precisely capture the state of gas at high redshifts. 

The horizontal dotted-dashed lines in each plots indicate CMB temperature ($T_{\rm{CMB}}$). Dashed lines indicate the expected gas temperature due to cosmic expansion, denoted as $T_{ada}$. At $z=30$, a considerable fraction of gas is heated above $T_{ada}$ in all the three codes. Meanwhile, little gas is heated above $T_{\rm{CMB}}$ in WIGEON and PLM-PM. In samples run with RAMSES, however, a considerable fraction of gas is hotter than $T_{\rm{CMB}}$. The efficiency of gas heating in RAMSES is remarkable, considering little gas has a density over two times of the cosmic mean and $T_{\rm{CMB}}$ is about six times of $T_{ada}$. Namely, the temperature of the heated gas increases more rapidly than $\rho^{2/3}$. On the other hand, a notable fraction of gas in the RAMSES is even colder than $T_{ada}$. The fraction of such gas in other two codes is much smaller. We will revisit and discuss these effects in section 3.2 and 3.3. 
 
The bottom panel of the Fig.~\ref{fig:vol_dis_ada} shows the evolution of cosmic gas in the radiative simulations with identical resolutions to the non-radiative runs shown above. Comparing to the non-radiative runs, radiative processes only have mild effects on the evolution of gas in the $(\rho, T)$ phase plane in the redshift range $30-11$. Therefore, gravitational collapse of structures and associated shocks dominate over radiative processes on the heating of gas in this era in our simulations. Considering the small effects of Compton and UV heating, and radiative processes, we will focus on the evolution of gas in non-radiative simulations in the following context of this subsection.  

\begin{figure*}
\begin{center}
\vspace{1.0cm}
\includegraphics[trim = 0mm 10mm 0mm 10mm, width=0.9\textwidth]{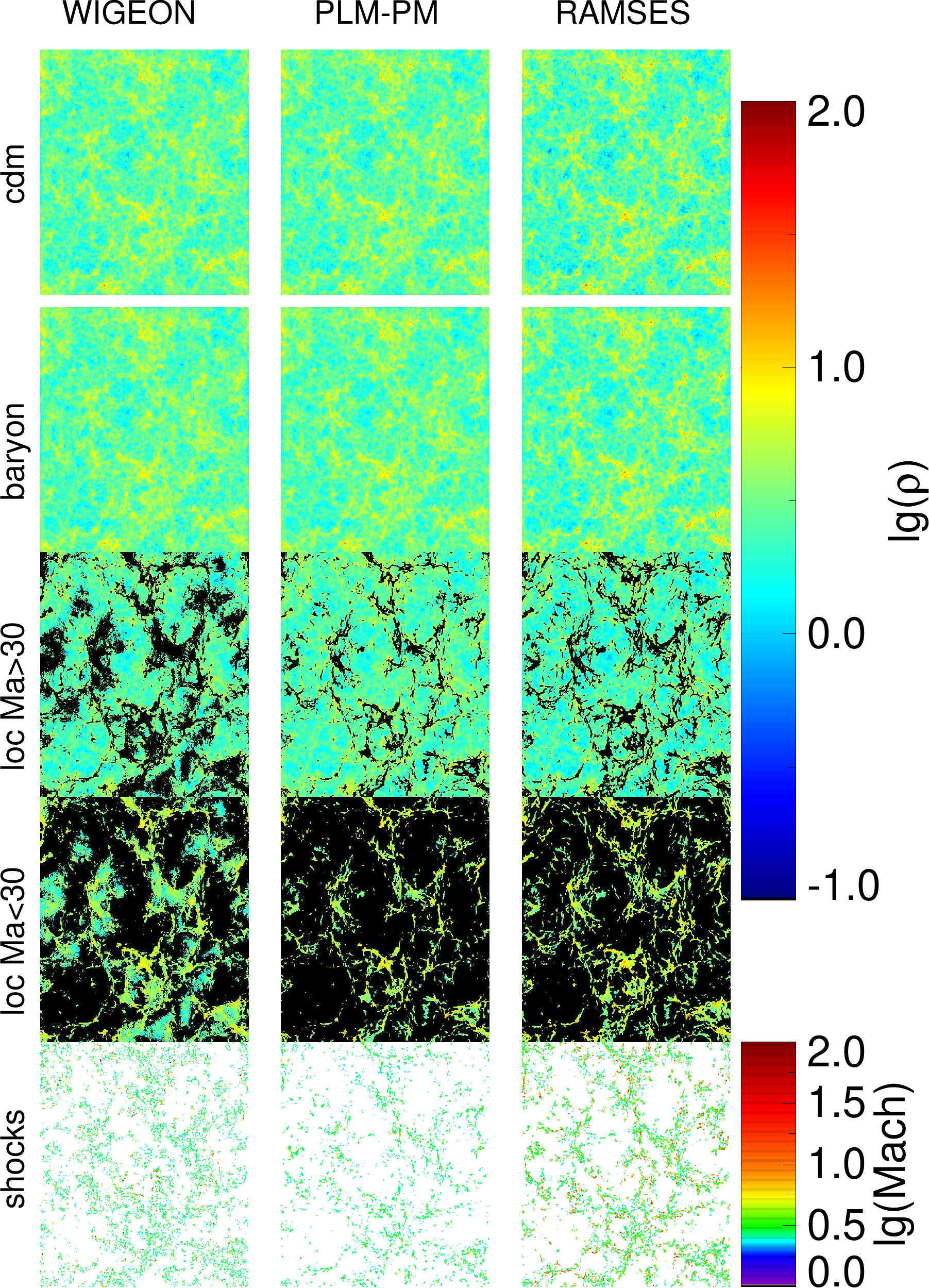}
\vspace{1.0cm}
\caption {Dark matter and gas density and shocks in a slice of depth $0.25 h^{-1}$ Mpc at $z=11$ in our non-radiative simulations. First and second row show dark matter density and gas density respectively, third and fourth row show the density of gas with $\rm{Ma}_{\rm{loc}}>30$ and $\rm{Ma}_{\rm{loc}}<30$ respectively. The fifth row presents locations and mach numbers of shocks. For the sake of clarity, only shocks with a mach number larger than 2.0 are shown.}
\label{fig:den_shock_slice}
\end{center}
\end{figure*}

Physically, the gas temperature in non-radiative simulations is largely determined by gravitational collapse heating and expansion cooling. It is natural to expect that the thermal evolution of gas residing in over-dense environment should differ from the gas in under-dense regions. The former might be influenced more by gravitational collapse. The thermal evolution of gas in these two regions may provide more insightful information on gravitational collapse heating at this era. We separate these two gas components using the local mach number, $\rm{Ma}_{\rm{loc}}=v/v_{cs}$. The local mach number of gas in over-dense regions is likely to be smaller due to enhanced internal energy by collapse heating. For gas in void region, on the other hand, the local mach number will be relatively large due to the low sound speed as a result of cooling, and considerable flow velocity. We select a threshold value of $\rm{Ma}_{\rm{loc}}=30$, i.e., the kinetic energy is about 500 times of the internal energy.

Fig.~\ref{fig:den_shock_slice} provides a visual impression of the distribution of the gas with a local mach number larger or smaller than 30. The density of dark matter and gas within a slice of depth $0.25 h^{-1}$ Mpc at redshift $z=11$ is presented in the first and second row of Fig.~\ref{fig:den_shock_slice}. The third and fourth row show the density of gas with $\rm{Ma}_{\rm{loc}}>30$ and $\rm{Ma}_{\rm{loc}}<30$, respectively. Apparently, gas residing in under-dense(void) region tends to have large local mach number. In contrast, gas with relatively small local mach number is more likely to be associated to high density peaks. 

We also show the distribution of shocks in this slice in the fifth row of Fig.~\ref{fig:den_shock_slice}. Shocks in our simulations are identified with the conventional method introduced in \citet{Miniati2000}. The divergence of velocity field and changes in density and temperature are used to locate the shock center. The mach number, $M$, of shocks is determined by the temperature change across the shocks. For the sake of clarity, only shocks stronger than $M>2.0$ are shown. At $z=11$, numerous shocks can be observed in this slice. Moreover, strong shocks are generally found in over-dense region. Hence, shocks should have played an important role in heating the gas in over-dense region, in consistent with expectation. The number of strong shocks in the WIGOEN and RAMSES run are much larger than that in PLM-PM. On the other hand, the simulations run with RAMSES contain more shocks with mach number larger than 5. 

\begin{figure}
\begin{center}
\vspace{-0.5cm}
\includegraphics[trim = 0mm 00mm 0mm 00mm, width=0.5\columnwidth]{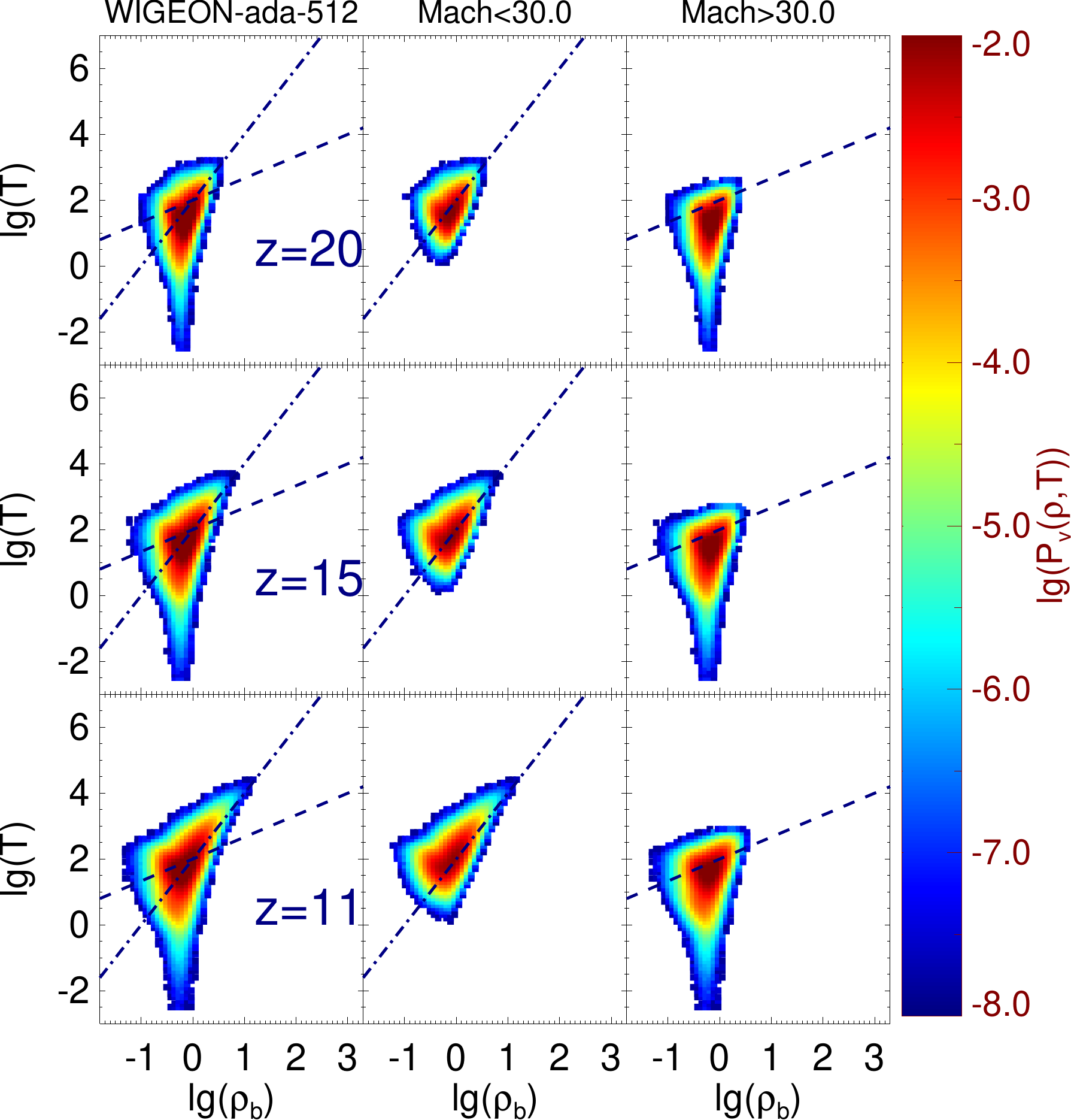}
\includegraphics[trim = 0mm 00mm 0mm 00mm, width=0.5\columnwidth]{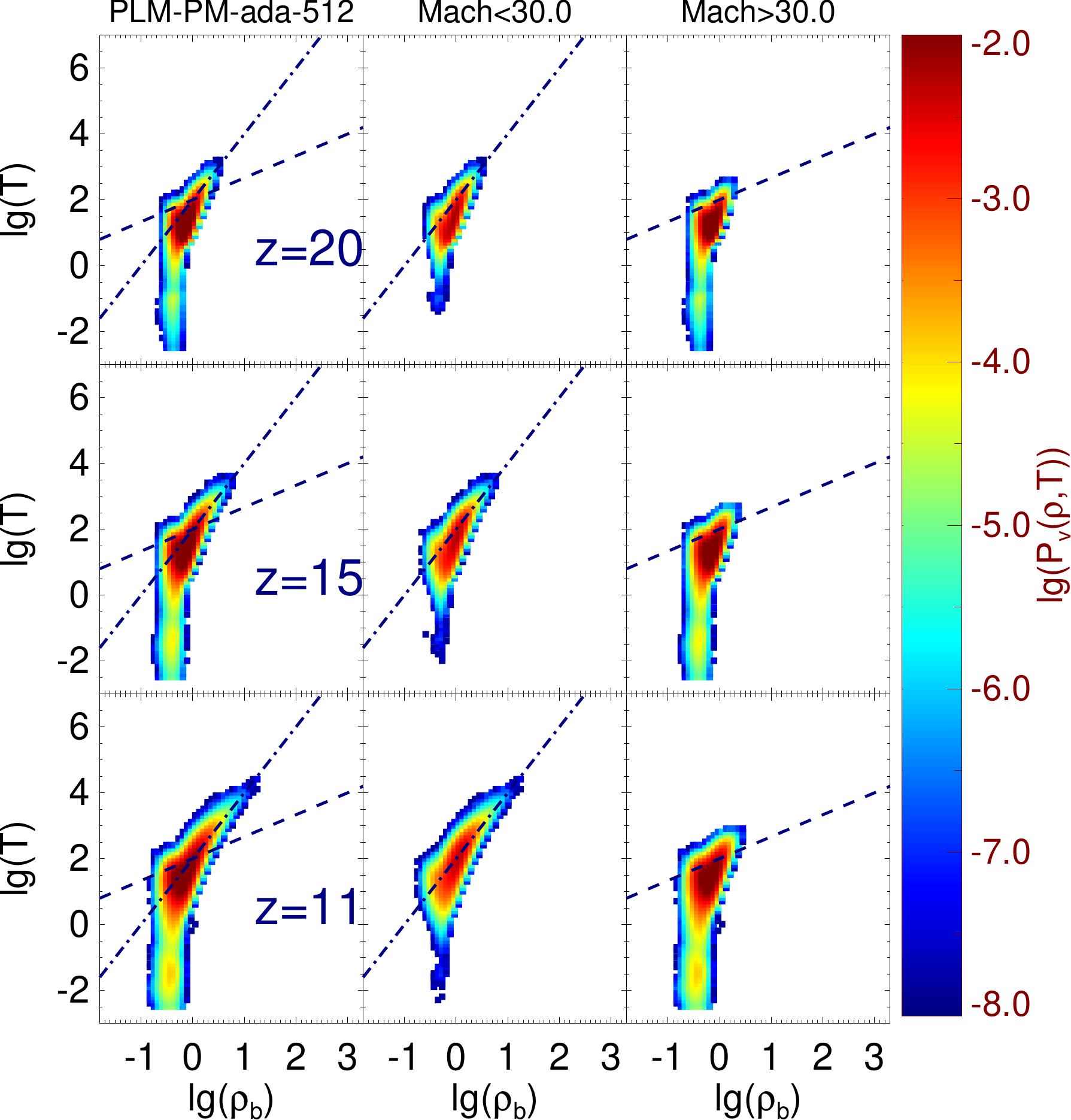}
\includegraphics[trim = 0mm 00mm 0mm 00mm, width=0.5\columnwidth]{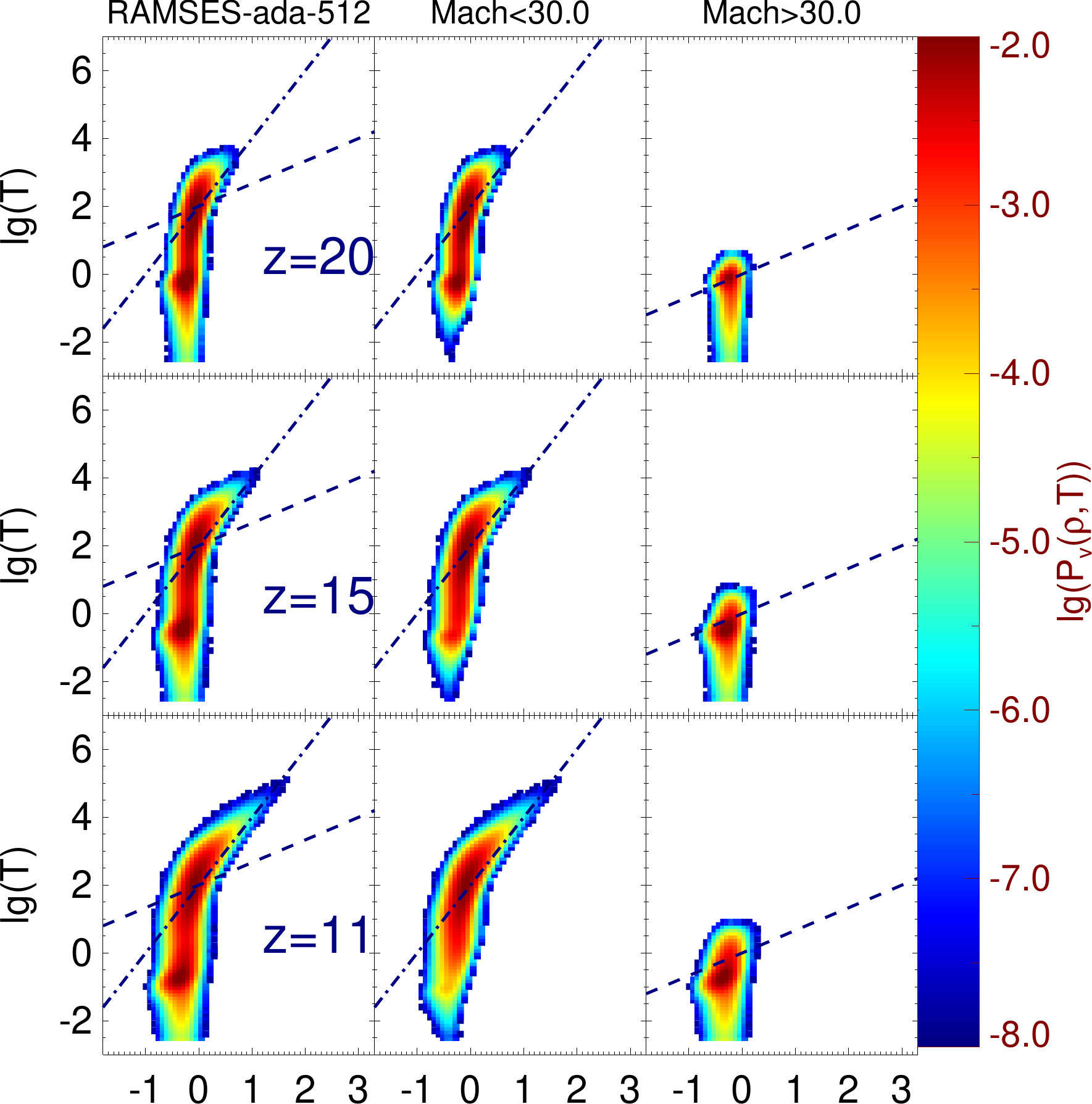}
\vspace{-0.2cm}
\caption{The distribution of the cosmic gas in the density-temperature plane at redshift $20, 15, 11$ in our non-radiative simulations. The left, middle and right column in each panel show all gas, the gas with local mach number below  and above 30, respectively. Dashed and dotted-dashed lines indicate $T \propto \rho^{2/3}$ and $T \propto \rho^{2}$ respectively. Number of bins in each plot have been reduced with respect to Fig.~\ref{fig:vol_dis_ada} in order to cut the size of figure.}
\label{fig:locma}
\end{center}
\end{figure}

We show the evolution of those two components of gas in the density-temperature phase diagram in Fig.~\ref{fig:locma}. In accordance with expectation, the thermal evolution of those two components are quite different. For gas with $\rm{Ma}_{\rm{loc}}<30$, there is a weak correlation between density and temperature at $z=20$. Nevertheless, the degree of correlation increases as redshift decreases. When density grows, the temperature of gas with $\rm{Ma}_{\rm{loc}}<30$ shows a tendency to increase more rapidly than $\rho^{2/3}$, i.e., the isentropic relation for ideal gas with $\gamma=\frac{5}{3}$. A crude approximation gives $T \propto \rho^{2}$.

For gas with $\rm{Ma}_{\rm{loc}}>30$, there is very weak sign of correlation between density and temperature in all our simulations. These gas generally have density $0.1-1.0$ times of the cosmic mean baryonic density. Most of the gas with $\rm{Ma}_{\rm{loc}}>30$ have temperatures in the range of $1-100$K in WIGEON-512-ada and PLM-PM-512-ada, and of $0.1-10$K in RAMSES-512-ada. Modest difference can be found between WIGEON and PLM-PM for gas with $\rm{Ma}_{\rm{loc}}>30$, which might be attributed to the different schemes used to solve the hydrodynamic equations and the temporal discretization between two codes. Distinctive discrepancy between samples run with RAMSES and other two codes may be partly caused by the different potential solvers, and the treatment of gravity term in fluid equations.  

For a short summary of this subsection, the cosmic gas has become multi-phase since a time earlier than $z=30$. Considerable differences on gas phases are observed among simulations run with different codes. In the redshift range $30-11$, the Compton heating and cooling, UV heating, and radiative cooling have negligible impact on the evolution of gas. The temperature of gas is largely determined by collapse heating, and cosmic expansion cooling. A notable fraction of gas is heated above the temperature of CMB photons at $z\sim 15 $. The evolution of gas temperature in and surrounding high density peaks is closely related to structure formation shocks, and shows a tendency to increase more sharply than $\rho^{2/3}$ as density grows, approximately $\rho^{2}$. At redshift $z \gtrsim 11$, the internal energy of gas in the under-dense region tends to be much smaller than the kinetic energy, i.e. have large local mach number. 

\subsection{thermal history of IGM at high redshifts: quantitative results}
\begin{figure}
\begin{center}
\vspace{0.0cm}
\includegraphics[trim = 0mm 0mm 0mm 0mm, width=0.48\columnwidth]{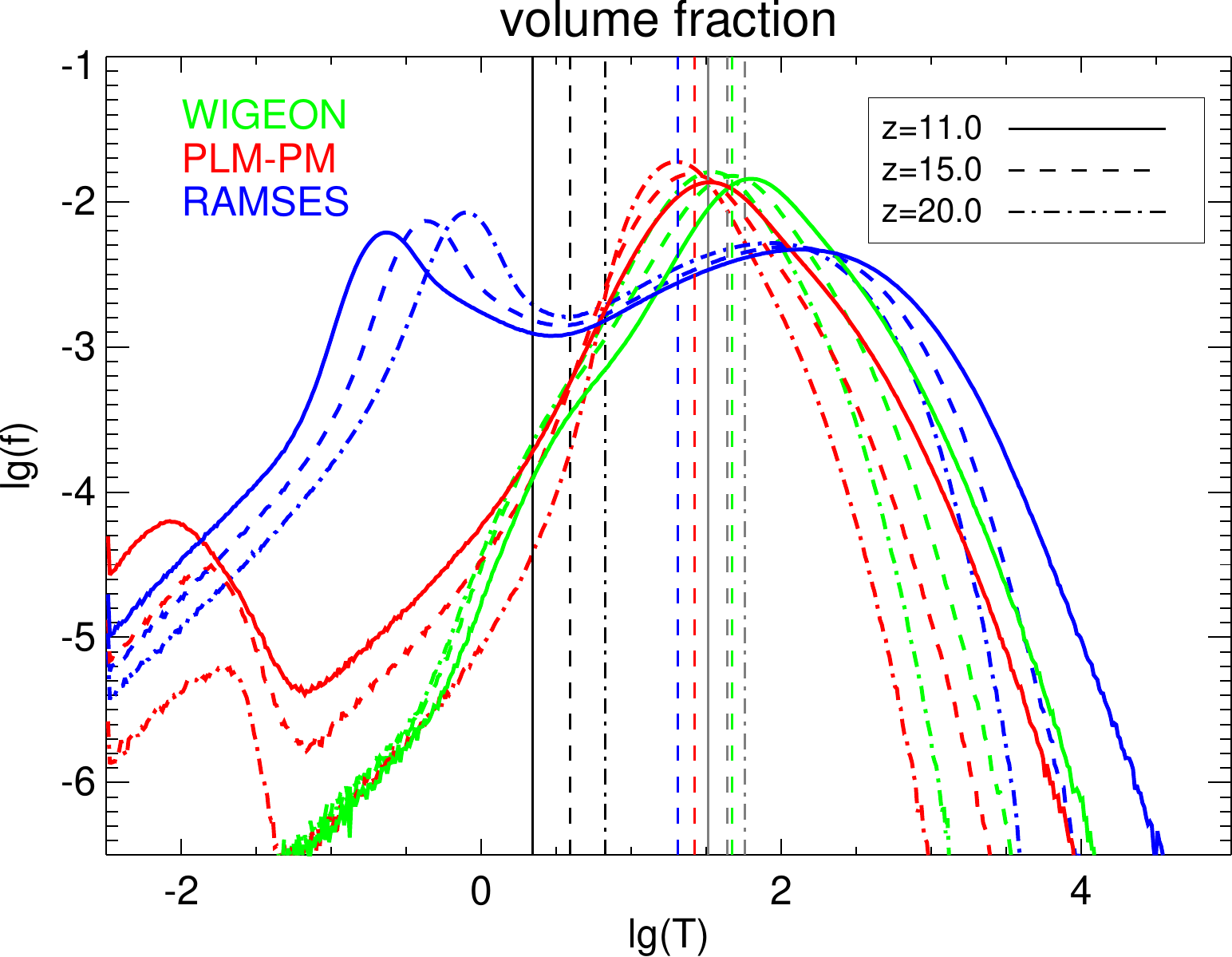}
\includegraphics[trim = 0mm 0mm 0mm 0mm, width=0.48 \columnwidth]
{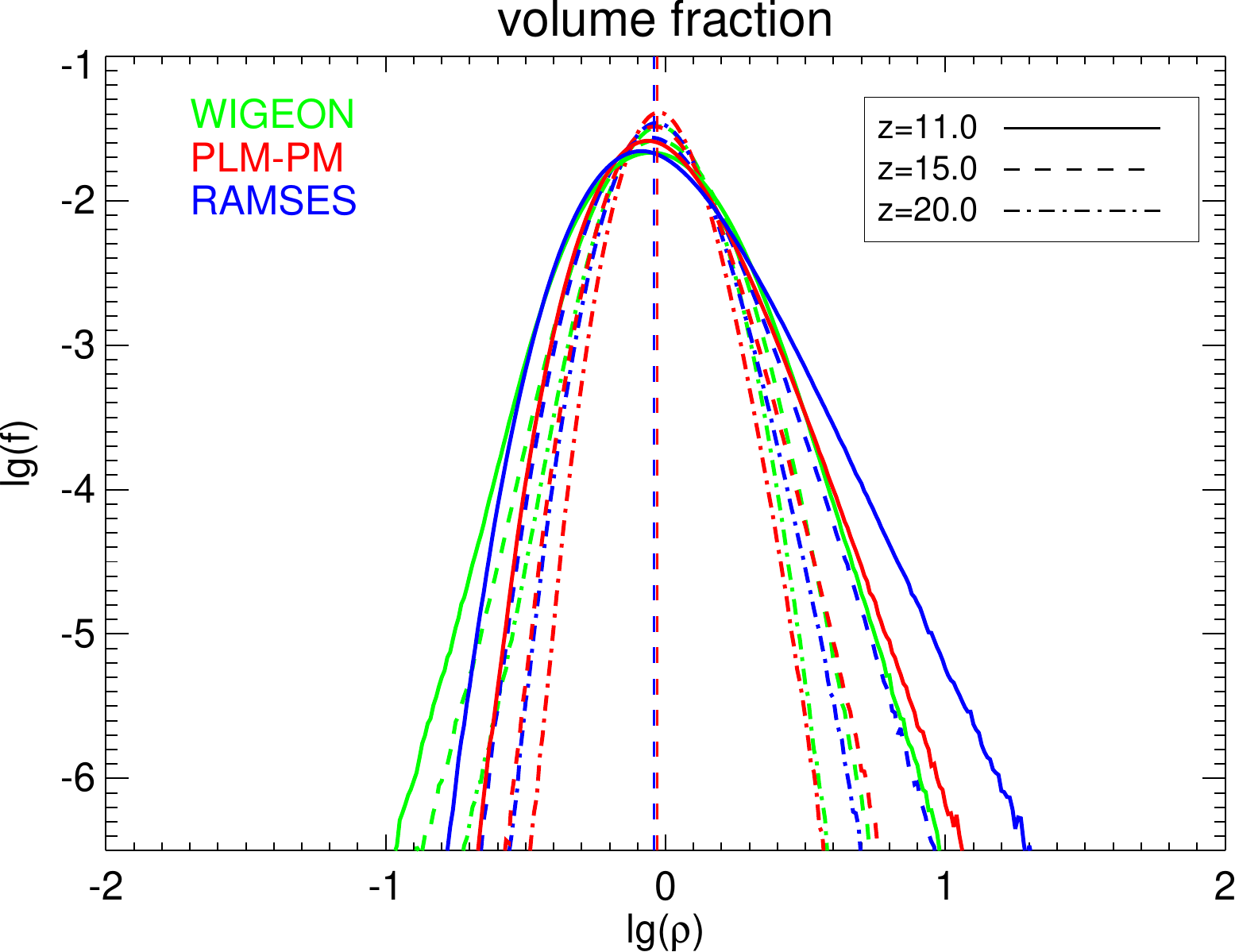}
\includegraphics[trim = 0mm 0mm 0mm 0mm, width=0.48\columnwidth]{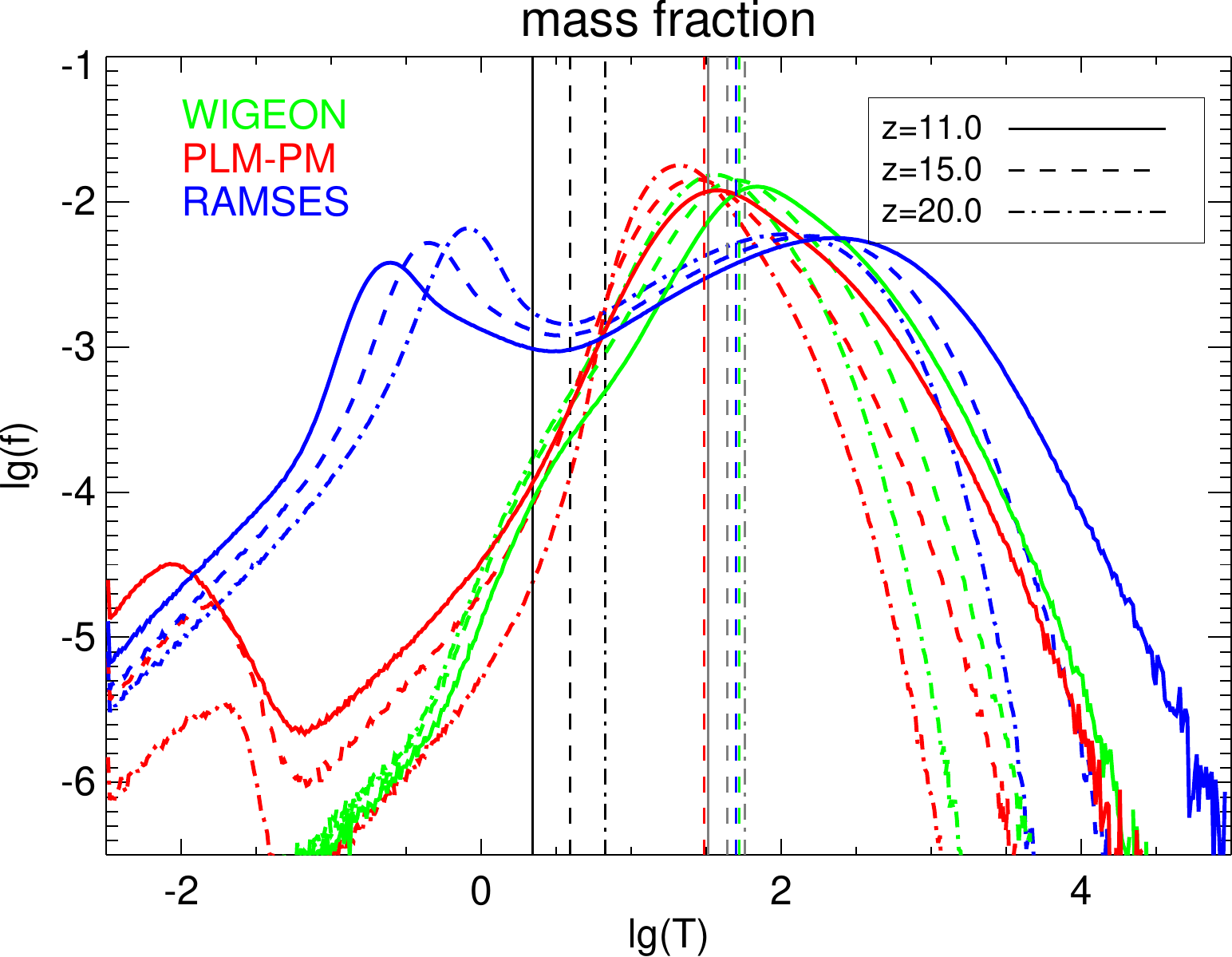}
\includegraphics[trim = 0mm 0mm 0mm 0mm, width=0.48\columnwidth]
{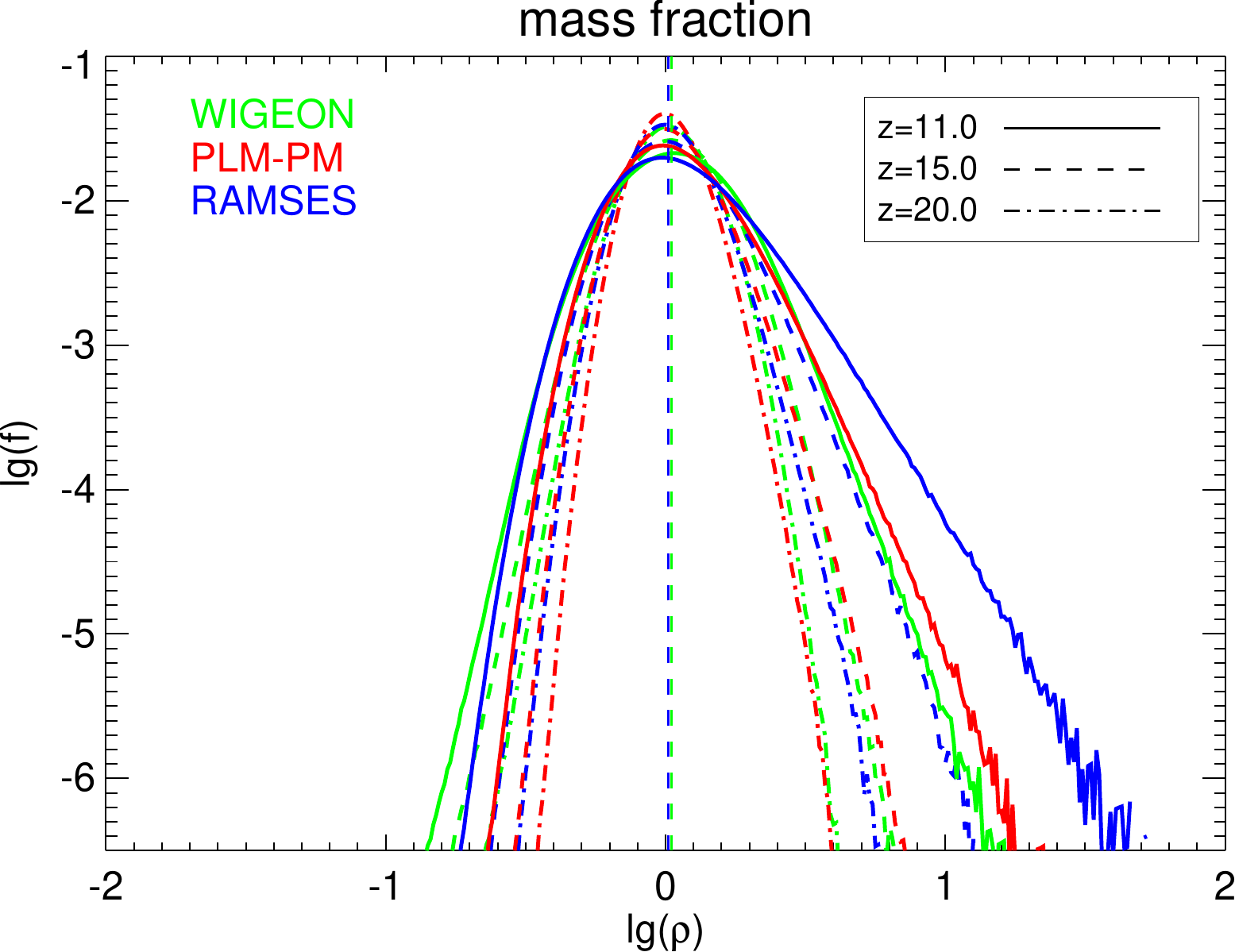}
\vspace{0.0cm}
\caption {Left column: Top(bottom) plot indicates distribution of volume (mass) fraction as a function of temperature in non-radiative simulations at redshift $z=20, 15, 11$. Vertical black and grey lines indicate $T_{ada}$ and $T_{\rm{CMB}}$ respectively, vertical colorful lines indicate median temperature for different codes at redshift 15. Right column: Top(bottom) plots indicates distribution of volume(mass) fraction as function of density in non-radiative simulaitions, vertical dash line represent median density.}
\label{fig:tem_hist}
\end{center}
\end{figure}

The thermal state of IGM are quite different among different codes. To assess the differences, we provide quantitative analysis in this subsection. In the left column of Fig.~\ref{fig:tem_hist} we show the distributions of volume and mass fraction of gas as a function of temperature in our non-radiative simulations at $z=20,15,11$. The vertical black and grey lines indicate $T_{ada}$ and $T_{\rm{CMB}}$  at corresponding redshifts, vertical colorful lines indicate median temperature for three codes at redshift 15. A bimodal distribution is found for the sample run with RAMSES. The peak at $T<1$K is associated with gas with a large local mach number and resides in under-dense region. The other peak temperature is higher than $T_{\rm{CMB}}$ at $z \lesssim 20$. The sample run with WIGEON exhibits a single peak distribution with the peak temperature close to $T_{\rm{CMB}}$. By and large, the distribution function in PLM-PM is similar to WIGEON with a slightly lower peak temperature. However, a bump at $T\sim 10^{-2}$K is observed for PLM-PM. In right column of Fig.~\ref{fig:tem_hist}, we show the distribution of volume and mass fraction of gas as function of density, vertical dashed lines represent the median density. Although RAMSES has slightly larger fraction of high density region, the three sets of simulations show a similar density distribution.

\begin{figure}
\begin{center}
\vspace{-0.55cm}
\includegraphics[trim = 0mm 0mm 0mm 0mm, width=0.6\columnwidth]{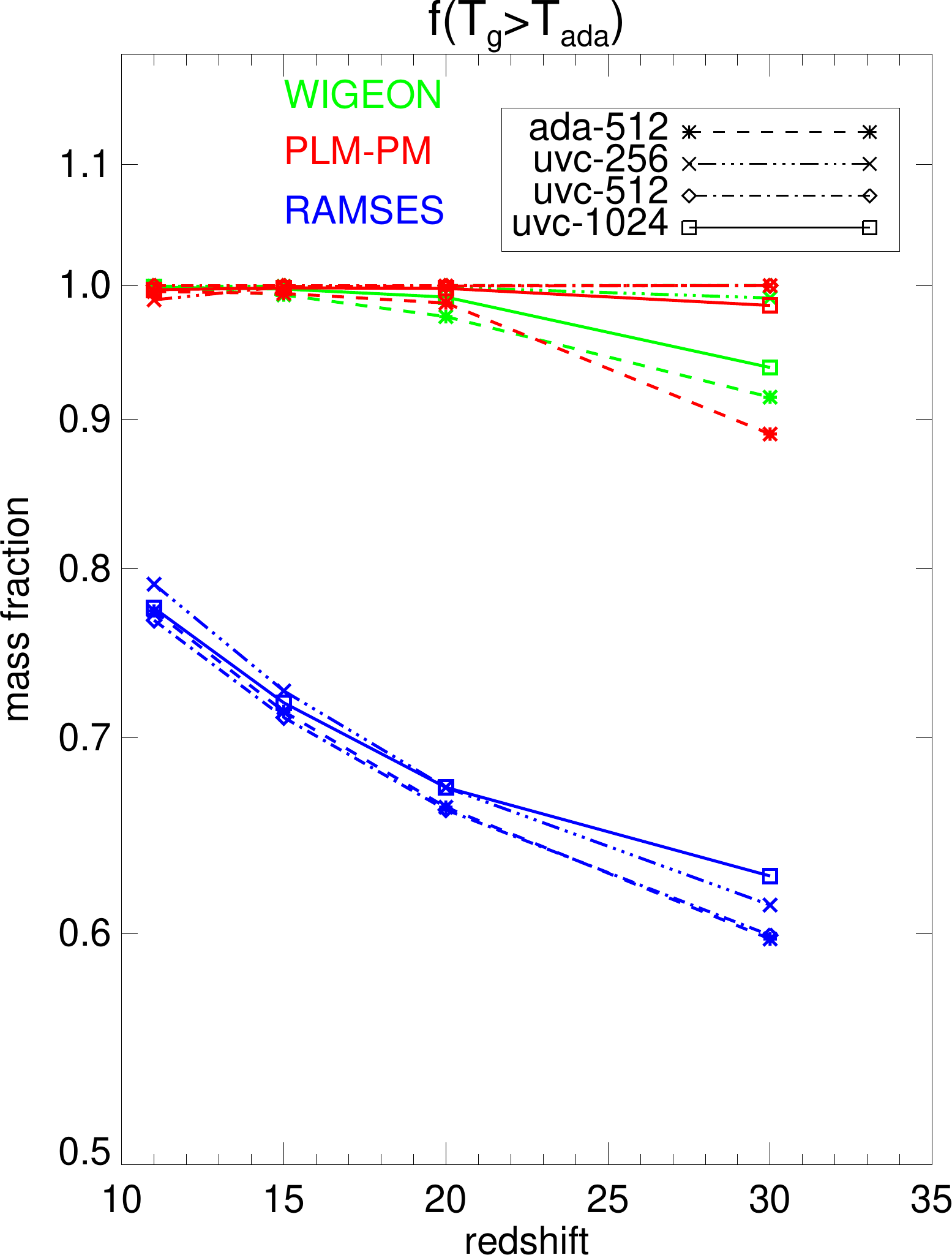}
\vspace{-0.25cm}
\includegraphics[trim = 0mm 0mm 0mm 0mm, width=0.6\columnwidth]{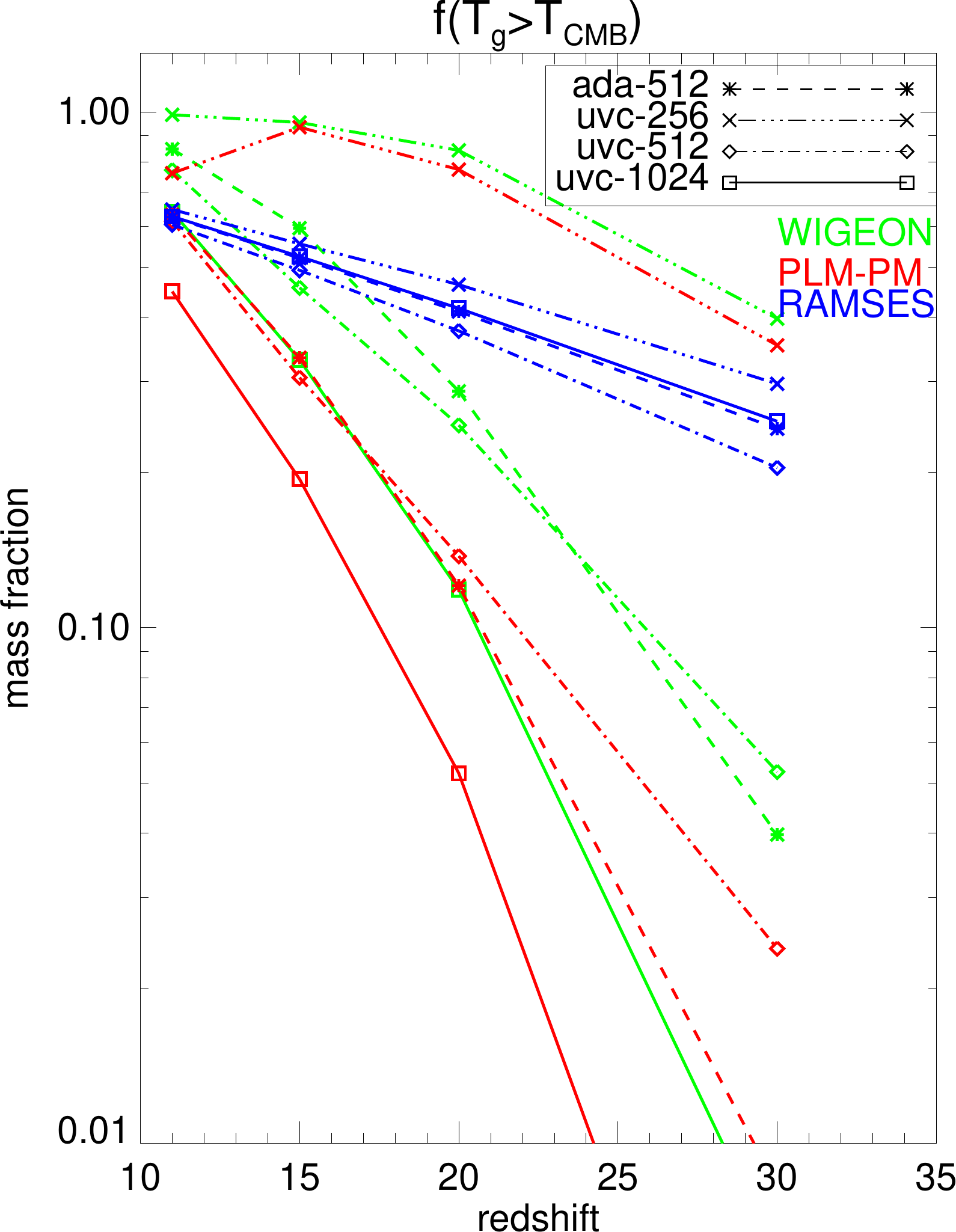}
\caption{Mass fractions of gas heated above $T_{ada}$ , and above CMB temperature $T_{cmb}$ between $z=30$ and $z=11$. Green, red and blue color indicate simulations run with WIGEON, PLM-PM and RAMSES, respectively. Asterisk indicates the non-radiative simulation with $512^3$ grids. {'x'}, diamond and square indicate the simulations including Compton process, UV background, and radiative cooling with $256^3$, $512^3$ and $1024^3$ grids respectively.}
\label{fig:massfrac}
\end{center}
\end{figure}

 We then measure the mass fraction of gas that has been heated above two characteristic values, $T_{\rm{CMB}}$ and $T_{ada}$. Fig.~\ref{fig:massfrac} shows the mass fraction above $T_{ada}$, denoted as $f(T_g>T_{ada})$, and above CMB temperature $T_{\rm{CMB}}$, denoted as $f(T_g>T_{\rm{CMB}})$, in the redshift range $30-11$. In the simulations run with WIGEON and PLM-PM, more than $90\%$ of the gas has been heated above $T_{ada}$ over the whole redshift range considered here. There is, however, about $25\%-35\%$ of the gas colder than $T_{ada}$ in the RAMSES simulation. Such gas has large local Mach number and can be as cool as $T<1$ K as demonstrated in previous figures. Comparing radiative and non-radiative simulations with the same resolution, we find that the physical processes included in this paper, i.e., Compton heating, UV heating and radiative heating and cooling, have minor influence on the mass fractions of gas with $T_g>T_{ada}$ in all our simulations, and on the fractions of gas with $T<1$ K in the RAMSES simulation. The increase in numerical resolution also has minor effect on $f(T_g>T_{ada})$ in all three codes. 

\begin{figure}
\begin{center}
\vspace{-0.3cm}
\includegraphics[trim = 0mm 0mm 0mm 0mm, width=0.7\columnwidth]{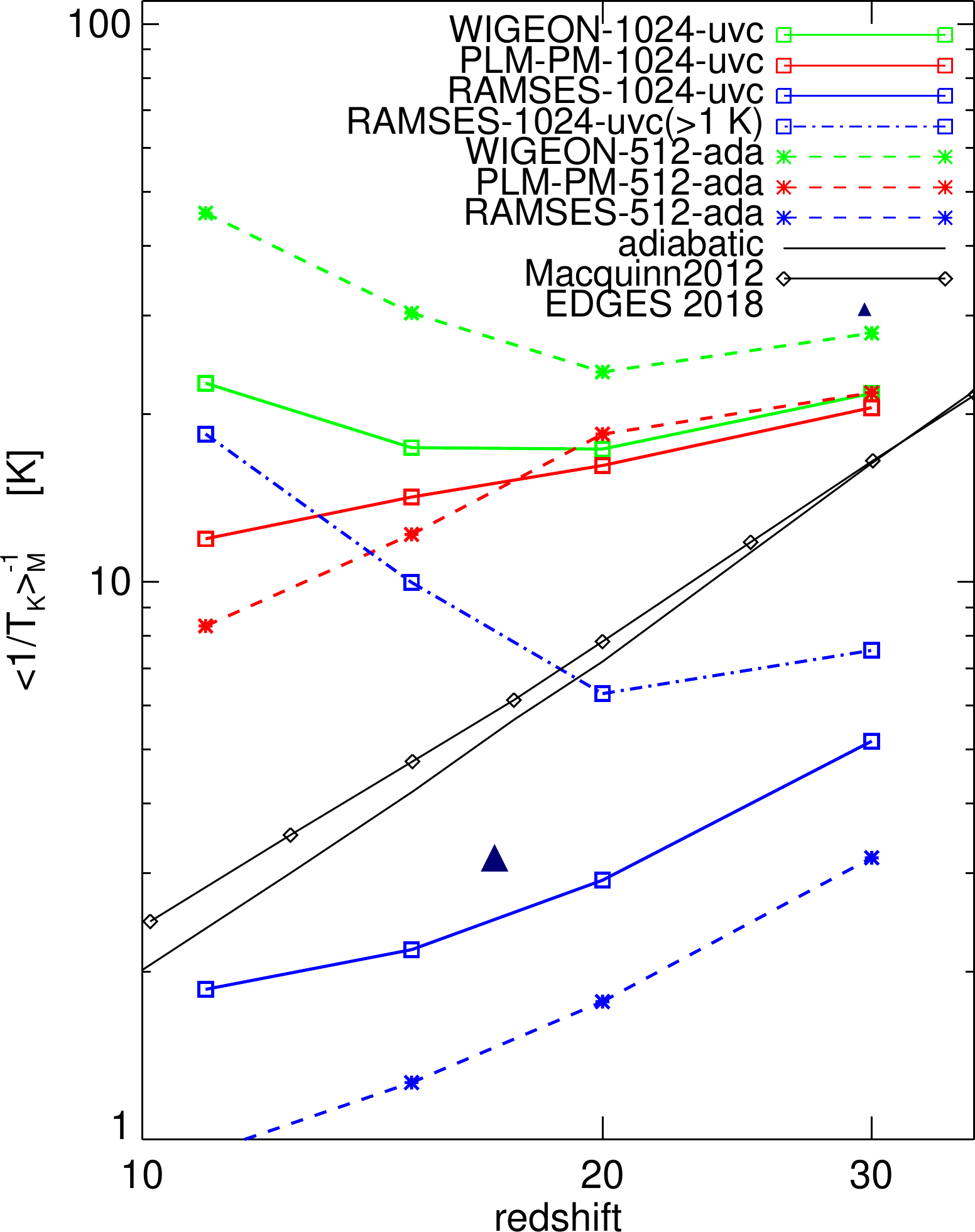}
\caption{Evolution of gas temperature $<1/T_{\rm{K}}>_M^{-1}$ in our highest resolution simulations. Dash-doted blue line indicates $<1/T_{\rm{K}}>_M^{-1}$ for the gas with $T>1$ K in RAMSES-uvc-1024. Dashed lines indicate $<1/T_{\rm{K}}>_M^{-1}$ in non-radiative simulations. Results in \citet{McQuinn2012} based on simulations run with Gadget and Enzo are  shown as open diamonds. Asterisks indicate the gas temperature expected from the Cosmic expansion. The filled triangle marks the result suggested by EDGES(\citealt{Bowman2018})}
\label{fig:tem_m}
\end{center}
\end{figure}

The mass fraction of gas above $T_{\rm{CMB}}$ increases as redshift decreases in the redshift range $11<z<30$. The fraction in the samples run with WIGEON and PLM-PM grows more rapidly than that in the samples run with RAMSES. At $z=30$, the fraction of gas with $T_g>T_{\rm{CMB}}$ varies significantly from code to code. In the samples run with RAMSES, the fraction can be as high as $\sim 25\%$ and is independent on numerical resolution and physical processes. However this fraction is only $\sim 1\%-5\%$ in the simulations run with WIGEON and PLM-PM with the highest resolution, but is $\sim 40\%$ in the simulations with a number of grids $256^3$. The discrepancies between RAMSES and other two codes narrow down gradually as redshift decreases.

At $z=20$, the mass fraction of gas with $T_g> T_{\rm{CMB}}$ ranges from $5\%$ to $40 \%$ in different simulations. Specifically, the fractions in the highest resolution simulations are $\sim 40 \%$, $\sim 12 \%$ and $\sim 5 \%$ for code RAMSES, WIGEON, PLM-PM respectively. Those fractions of gas with $T_g> T_{\rm{CMB}}$ grow to $\sim 52 \%$, $\sim 33 \%$, and $\sim 19 \%$ at $z=15$, and further grow to $\sim 63 \%$, $\sim 64 \%$ and $\sim 44 \%$ at $z=11$. The Compton process, UV, and cooling have minor impact on the results of all the three codes at spacial resolution $48.8 h^{-1}$ kpc. In simulations including multi-physical processes, the mass fraction of gas above $T_{\rm{CMB}}$ decreases with increasing resolution for the simulations run with WIGENON and PLM-PM but not for the ones run with RAMSES.

In the literature, \citet{Furlanetto2006} estimated that at redshifts $30, 20, 10$, about $0.1 \%, 3\%, 25\%$ of the cosmic gas should be shock heated to $T_g> T_{\rm{CMB}}$, respectively. \citet{McQuinn2012} found that the temperature deviations in their simulations were consistent with \citet{Furlanetto2006}, but about half of the gas heating was due to the Compton heating. However, the impact of shock heating was much more significant in \citet{Gnedin2004}. The mass fractions of gas with $T_g> T_{\rm{CMB}}$ in our simulations are much higher than the results of \citet{Furlanetto2006}, by a factor of 10-50 at $z=30$, and by a factor of 2 to 8 in the redshift range $11-20$. In general, our results are more close to \citet{Gnedin2004}.

We further show the mass averaged $<1/T_{\rm{K}}>_M^{-1}$ in our highest resolution simulations in Fig.~\ref{fig:tem_m}. Here $<1/T_{\rm{K}}>_M^{-1}$ is the gas temperature most relevant to the HI 21 cm signals(see the Appendix A in \citealt{McQuinn2012}). For the samples run with WIGEON and PLM-PM, $ <1/T_{\rm{K}}>_M^{-1}$ is significantly higher than the expectation from adiabatic expansion, as well as the results in \citealt{McQuinn2012}). While, $<1/T_{\rm{K}}>_M^{-1}$ in RAMSES-uvc-1024 is well below the value expected from adiabatic evolution in the redshift range $30-11$. It is even colder than the gas temperature at $z=17$ suggested by EDGES(\citealt{Bowman2018}). This is because a significant fraction of gas in the simulations run with RAMSES has temperature as cold as $\sim 1$ K. When excluding this part of gas, $<1/T_{\rm{K}}>_M^{-1}$ in RAMSES-uvc-1024 is larger than $T_{ada}$ at $z<20$ and is more close to the results derived by other two codes.

\subsection{Shocks at high redshifts}

One of the most significant differences among the simulations presented here, as well as in the literature, is the value of $f(T_g>T_{\rm{CMB}})$ at $Z>10$. This discrepancy was mainly attributed to different resulting shock heating in different simulations(\citealt{Gnedin2004, McQuinn2012}). However, it is not easy to quantify the number and intensity of shocks precisely with analytical method. Furthermore, the statistical properties of shocks at $z>10$ are not provided in previous simulation works. Fig.~\ref{fig:den_shock_slice} indicates that the number of strong shocks found in our simulations indeed varies from code to code.

A comparison on the number and intensity of identified shocks between simulations can help to assess the impact of shock heating and understand the cause of discrepancy on $f(T_g>T_{\rm{CMB}})$. Here, we study the frequency of shocks found in our non-radiative simulations run with different codes, and its redshift evolution. We specifically examine non-radiative simulations because the differences in these samples are not relevant with radiative processes. In addition, the differences among codes for non-radiative simulations are comparable to radiative simulations with the same resolution.

\begin{figure}
\begin{center}
\vspace{0.0cm}
\includegraphics[trim = 0mm 0mm 0mm 0mm, width=0.7\columnwidth]{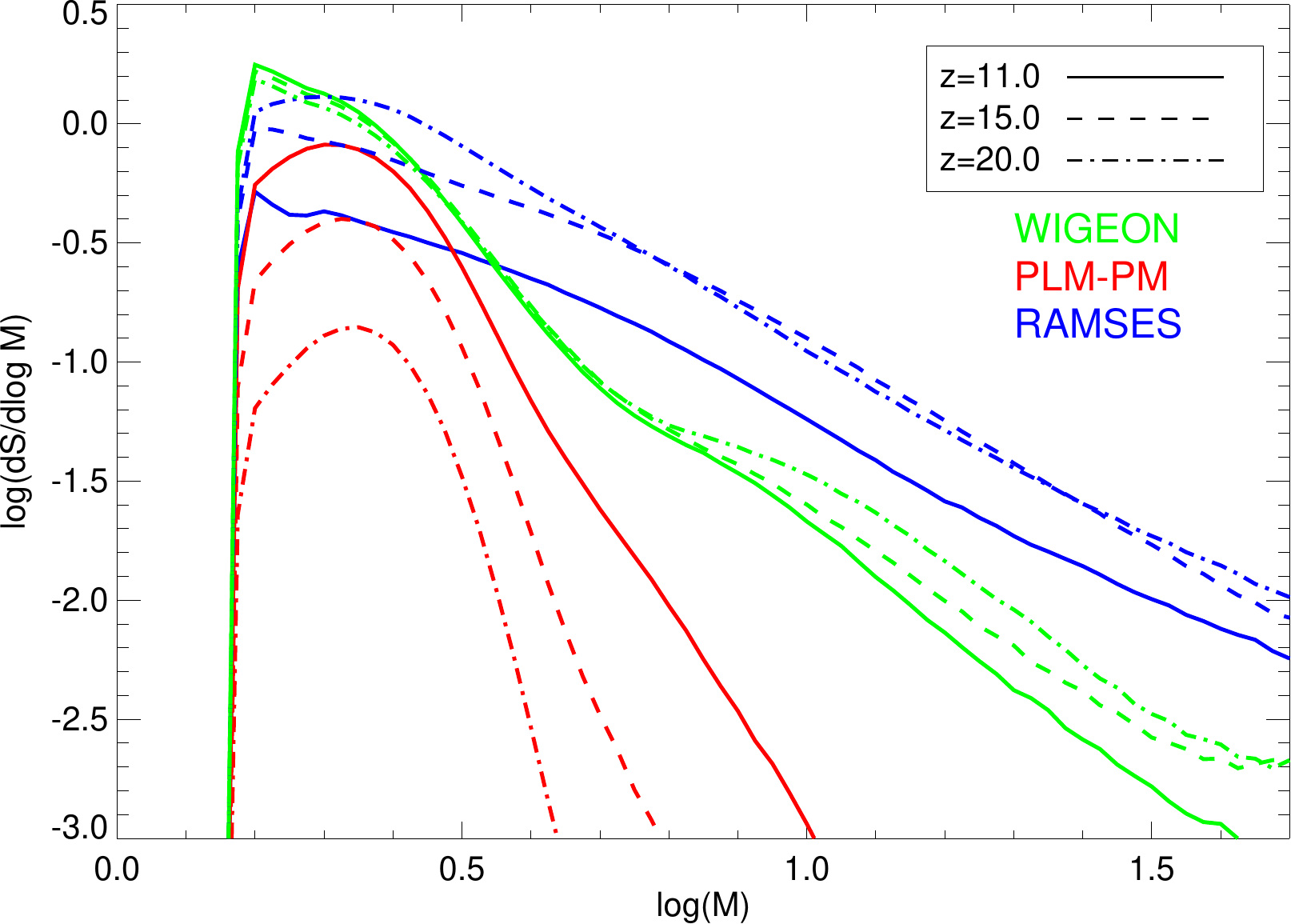}
\vspace{0.0cm}
\caption{Surface area of shocks as a function of mach number in our non-radiative simulations at redshift $z=20, 15, 11$. }
\label{fig:shock_freq}
\end{center}
\end{figure}

The frequency of shocks can be measured with the surface area of shocks, $d S(M, z)/ d log M$, which reveals the inverse distance between shocks(\citealt{Miniati2000}). Fig.~\ref{fig:shock_freq} shows the surface area of shocks identified in our non-radiative simulations at redshift $z=20, 15, 11$. At $z=20$, samples run with RAMSES contain the biggest number of shocks, while PLM-PM contain the smallest number. As redshift decreases, the number of shocks decreases slightly in the WIGEON-ada-512, and moderately in the RAMSES-ada-512, but increases sharply in the PLM-PM-ada-512. At $z=11$, the simulation run with WIGEON comprises more shocks with mach number smaller than $\sim 3.5$ than the samples run with the other two codes. Meanwhile, shocks with mach number $\rm{M}> \sim 3.5$ are more ready to be found in the samples run with RAMSES. The total frequency of shocks is comparable between the samples run with WIGEON and RAMSES, but is the lowest for the PLM-PM simulation, suggesting that the WENO scheme has better ability to captures shocks than the PLM scheme. On the other hand, gravity solver also plays an important role in developing shocks at high redshifts, which is reflected by the difference in the frequency of shocks between the PLM-PM and  RAMSES simulations. These two codes use the same hydrodynamical scheme, but adopt different gravity solvers. Thus, both the hydrodynamical and gravity solver are important to capture the structure formation shocks at high redshifts.

The evolution of shocks identified in different simulations is consistent with the evolution of gas phases and corresponding quantitative results presented in the last two subsections. At $z=20$, the fraction of gas with $T_g>T_{\rm{CMB}}$ is considerable in samples run with RAMSES, and much larger than WIGEON and PLM-PM. This fits with the frequency of shocks found in various samples at that epoch. As redshift decreases, the discrepancy on $f(T_g>T_{\rm{CMB}})$ in three codes narrows down gradually, which agrees with the result on change of shock frequency.

On the other hand, as Fig.~\ref{fig:locma} shows that the gas demonstrates a tendency to evolve faster than $T \propto \rho^{2/3}$ in over-dense region, approximately evolve as $\rho^2$. This tendency is also likely to be related to shocks. The density and temperature of gas flow in the pre-shock and post-shock regions fulfill(\citealt{Landau1959}) 

\begin{equation}
\frac{\rho_2}{\rho_1}=\frac{4M^2}{M^2+3},
\end{equation}
\begin{equation}
\frac{T_2}{T_1}=\frac{(5M^2-1)(M^2+3)}{16M^2},
\end{equation}

where $M$ is the mach number of shock, and the subscripts 1 and 2 indicate the pre-shock and post-shock regions, respectively. For a strong shock with mach number $M=2, 3,$ and 6, $T_2/T_1=2.08, 3.67,$ and $12.12$, i.e., $\approx (\rho_2/\rho_1)^{0.88}$, $(\rho_2/\rho_1)^{1.18}$ and $(\rho_2/\rho_1)^{1.92}$ respectively. A considerable number of shocks stronger than $M=2 \sim 3$ are captured in over-dense region in all the simulations, which may drive the gas temperature to increase more rapidly than $\rho^{2/3}$ when flowing into density peaks. Shocks stronger than $M=6$ are also captured, especially in the simulations run with RAMSES and WIGEON. These strong shocks should have played primary a role in heating gas above $T_{\rm CMB}$.

\section{Discussions and Conclusions}
The thermal history of cosmic gas in the Dark Ages remains largely unknown. Recently, the EDGES team reported a strong 21 cm absorption signal in the redshift range of $15-20$, which suggests the gas might be as cold as $\sim 3.2$ k at that time. To provide an accurate model to explain this observation and future detection, it is important to quantify the impact of relevant physics on the thermal evolution of gas between $z=10$ and $z \sim 30$. In this work, we have revisited the heating of gas by structure formation shocks by using a suit of fixed grid cosmological hydrodynamical simulations run with different numerical codes. We explore the thermal history of the gas in density-temperature phase plane, fractions of gas witch has been heated significantly, the spacial distribution and frequency of shocks in the redshift range of $11-30$. Our findings can be summarized as follows: 

1.The cosmic gas is in muti-phase state since a time earlier than $z=30$.  Numerous structure formations shocks emerge in the over-dense region since $z\gtrsim 20$, and are the primary physics to heat gas above the CMB temperature $T_{\rm{CMB}}$. The gas surrounding high density peaks gradually develops a relation more sharp than $T \propto \rho^{2/3}$, approximately $T \rho^{2}$ from $z=30$ to $z=11$. Meanwhile, the gas in under-dense region tends to have a large local mach number, and its thermal state shows significant discrepancies among simulations run with different codes. 

2. Both the hydrodynamical and gravity solvers in cosmological simulation code are very important to resolve the structure formation shocks at high redshifts. Difference in the numerical scheme in either hydrodynamical or gravity solver can lead to notable discrepancy in the number and strength of shocks, and therefore has significant impact on the thermal state of cosmic gas at $z \gtrsim11$ in simulations, e.g., the fraction of gas heated above $T_{\rm{CMB}}$ is quite different among different codes.

3. In the redshift range $11-20$, the mass fraction of gas heated above the CMB temperature in our simulations is larger than the estimation in \citet{Furlanetto2006} by a factor of 2 to 8. At $z=15$, the fractions varies from $\sim 19\%$ to $52 \%$ in simulations run with the different codes. Nevertheless, our results are more consistent with earlier simulation work in \citet{Gnedin2004}. The measured gas temperature $<1/T_{\rm{K}}>_M^{-1}$ in our WIDGEON and PLM-PM simulations is about $10-20$ K between $z=11$ and $z=20$, higher than  \citealt{McQuinn2012}). In simulations run with RAMSES,  $<1/T_{\rm{K}}>_M^{-1}$ is even below the temperature suggested by EDGES. This result is however biased by $20\%-30\%$ gas with extreme low temperature and large local mach number.

There are, however, some uncertainties in our results. Firstly, the numerical results do not always converge in our radiative simulations. For RAMSES simulations, the fraction of gas with $T>T_{\rm{CMB}}$ varies a few percent when the spacial resolution increases by a factor of 2. For WIGEON and PLM-PM runs, however, the fractions decrease by a absolute value of $\sim 10\%$. The difference in the resolution effect among different codes may be due to different gravity solvers adopted by different codes, as well as the way to tackle gravitational source terms in solving Euler equations numerically. Small errors in the calculation of kinetic energy and total energy can significantly change the temperature of gas with high mach number, which are very common in the dark ages. Consequently, the temperature of gas and the role of shock heating may be overestimated or underestimated due to such errors. Especially, the method we apply to identify shocks uses the gas temperature to estimate shock strength. 

Secondly, the Compton heating, UV background and radiative cooling included in our simulations have small impact on the temperature of gas. It is possible that these processes are not well resolved in our simulations with our limited resolution, as reflected by the fact that some of our results depend on resolution. In addition, star formation, molecular hydrogen formation and destruction, radiative transfer are not included in our simulations. While the former two modules are important in highly clustered regions, the latter may also have impact on the temperature of gas in under-dense region.

We conclude that, given the fact that predictions of the thermal history of cosmic gas in the dark ages diverge significantly among different numerical schemes, e.g. gravity solver, hydro solver, treatment of gravitational source terms, and time integration method, it seems a challenge to precisely describe the thermal evolution of cosmic gas in the redshift range $30-11$ in current cosmological hydrodynamical simulations. Further work on the numerical accuracy of Cosmological simulation codes is highly demanding.

\section*{Acknowledgements}
This work is supported by the National Key R\&D Program of China(NO. 2017YFB0203300), and the Key Program of the National Natural Science Foundation of China (NFSC) through grant 11733010. W.S.Z. acknowledges support from the NSFC grant 11673077. F.L.L. is supported by the NSFC grant 11333008 and the State Key Development Program for Basic Research of China (2015CB857000).





\begin{thebibliography}{99}

\bibitem[\protect\citeauthoryear{Barkana}{2018}]{Barkana2018}
Barkana R., 2018, Nature, 555, 71

\bibitem[\protect\citeauthoryear{Bowman, Rogers, \& Monsalve et al.}{2018}]{Bowman2018}
Bowman, J. D., Rogers, A. E. E.,  Monsalve, R. A.,  Mozdzen, T. J., \& Mahesh, N. 2018, Nature, 555, 67

\bibitem[\protect\citeauthoryear{Bregman}{2007}]{Bregman2007}
Bregman, J. N., 2007, \araa, 45, 221

\bibitem[\protect\citeauthoryear{Bryan, Norman \& Stone et al.}{1995}]{Bryan1995}
Bryan, G. L., Norman, M. L., Stone, J. M., Cen, R., \& Ostriker, J. P. 1995, Comput. Phys. Commun., 89, 149

\bibitem[\protect\citeauthoryear{Cen, Miralda-Escude, Ostriker et al.}{1994}]{Cen1994}
Cen, R. Y., Miralda-Escude, Jordi., Ostriker, J. P.,  Rauch, M. 1994, ApJ, 437L, 9

\bibitem[\protect\citeauthoryear{Cen \& Ostriker}{1999}]{Cen1999}
Cen, R. Y., Ostriker, J. P., 1999,  ApJ, 514, 1

\bibitem[\protect\citeauthoryear{Cui et.al.}{2019}]{Cui2019}
Cui W., et al., 2019, MNRAS, 485, 2367

\bibitem[\protect\citeauthoryear{Dave, Cen, Ostriker et al.}{2001}]{Dave2001}
Dave, R., Cen, R. Y., Ostriker, J. P., Bryan, G. L., Hernquist, L., et al. 2001, ApJ, 552, 473

\bibitem[\protect\citeauthoryear{Feng, \& Holder}{2018}]{Feng2018}
Feng C., Holder G., 2018, \apj, 858, L17

\bibitem[\protect\citeauthoryear{Feng, Shu \& Zhang}{2004}]{Feng2004}
Feng, L. L., Shu, C.-W., \& Zhang, M. P. 2004, ApJ, 612, 1

\bibitem[\protect\citeauthoryear{Furlanetto \& Loeb}{2004}]{Furlanetto2004}
Furlanetto, S. R., \& Loeb, A. 2004, \apj, 611, 642

\bibitem[\protect\citeauthoryear{Furlanetto, Oh, \& Briggs}{2006}]{Furlanetto2006}
Furlanetto, S. R., Oh, S. P., \& Briggs, F. H. 2006, Phys. Rep., 433, 181

\bibitem[\protect\citeauthoryear{Gnedin \& Shaver}{2004}]{Gnedin2004}
Gnedin, N. Y., \& Shaver, P. A. 2004, ApJ, 608, 611

\bibitem[\protect\citeauthoryear{Gunn \& Peterson}{1965}]{Gunn1965}
Gunn, J. E., Peterson, B. A. 1965. ApJ 142, 1633

\bibitem[\protect\citeauthoryear{Haardt \& Madau}{1996}]{Haardt1996}
Haardt, F., \& Madau, P.  1996, \apj, 461, 20

\bibitem[\protect\citeauthoryear{Hu, Kim, Cowie et al.}{1995}]{Hu1995}
Hu E. M., Kim T. S., Cowie L. L., Songaila A, Rauch M. 1995,  AJ, 110, 1526

\bibitem[\protect\citeauthoryear{Jiang \& Shu}{1996}]{Jiang1996}
Jiang, G., \& Shu, C.-W. 1996, Journal of Computational Physics, 126, 202

\bibitem[\protect\citeauthoryear{Landau \& Lifshitz}{1959}]{Landau1959}
Landau, L., \& Lifshitz, E. 1959, Fluid Mechanics (Oxford: Pergamon)

\bibitem[\protect\citeauthoryear{Madau, Meiksin, \& Rees}{1997}]{Madau1997} 
Madau, P., Meiksin, A., \& Rees, M. J., 1997, \apj, 475, 429

\bibitem[\protect\citeauthoryear{McQuinn}{2016}]{McQuinn2016}
McQuinn, M., 2016, ARAA, 54, 313

\bibitem[\protect\citeauthoryear{McQuinn \& O'Leary }{2012}]{McQuinn2012}
McQuinn, M., \& O'Leary, R. M.  2012, \apj, 760, 3

\bibitem[\protect\citeauthoryear{Miniati, Ryu \& Kang et al.}{2000}]{Miniati2000}
Miniati, F., Ryu, D., Kang, H., et al. 2000, ApJ, 542, 608

\bibitem[\protect\citeauthoryear{Nicastro}{2018}]{Nicastro2018}
Nicastro, F., Kaastra, J., Krongold, Y., et al. 2018, Nature, 558, 406

\bibitem[\protect\citeauthoryear{O'Leary \& McQuinn}{2012}]{OLeary2012}
O'Leary, R. M., \& McQuinn, M., 2012, \apj, 760, 4

\bibitem[\protect\citeauthoryear{Planck Collaboration et al.}{2014}]{Planck2014}
Planck Collaboration, Ade, P. A. R., Aghanim, N., et al. 2014, A\&A, 571, 16

\bibitem[\protect\citeauthoryear{Rauch, Miralda-Escude, Sargent, et al.}{1997}]{Rauch1997}
Rauch, M., Miralda-Escude, J., Sargent, W. L. W., Barlow, T. A., Weinberg D. H., et al. 1997, ApJ, 489, 7

\bibitem[\protect\citeauthoryear{Shull, Smith, Danforth}{2012}]{Shull2012}
Shull. J. M., Smith, B. D., Danforth, C. W. 2012,  ApJ, 759, 23

\bibitem[\protect\citeauthoryear{Sunyaev\& Zel'dovich}{1972}]{Sunyaev1972}
Sunyaev, R. A., \& Zel'dovich, Ya. B. 1972, A\&A, 20, 189

\bibitem[\protect\citeauthoryear{Theuns, Leonard, Efstathiou et al.}{1998}]{Theuns1998}
Theuns, T., Leonard, A., Efstathiou, G., Pearce, F. R., \& Thomas, P. A. 1998,
MNRAS, 301, 478

\bibitem[\protect\citeauthoryear{Tseliakhovich \& Hirata}{2010}]{Tseliakhovich2010}
Tseliakhovich, D., \& Hirata, C. 2010, Phys. Rev. D, 82, 083520

\bibitem[\protect\citeauthoryear{Toro}{1997}]{Toro1997}
Toro, E. F. 1997, Riemann Solvers and Numerical Methods for Fluid Dynamics (Berlin: Springer)

\bibitem[\protect\citeauthoryear{Teyssier}{2002}]{Teyssier2002}
Teyssier, R., 2002, A\&A, 385, 337

\bibitem[\protect\citeauthoryear{Zhang \& Shu}{2012}]{Zhang2012}
Zhang, X., \& Shu, C.-W. 2012, Journal of Computational Physics, 231, 2245

\bibitem[\protect\citeauthoryear{Zhu, Feng, Xia et al.}{2013}]{Zhu2013}
Zhu, W. S., Feng, L. L., Xia, Y. H., et al. 2013, ApJ, 777, 48
\end{thebibliography}




\appendix

\section{Density power spectrum}
\begin{figure}
\begin{center}
\vspace{-0.5cm}
\includegraphics[trim = 0mm 0mm 0mm 0mm, width=0.85\columnwidth]{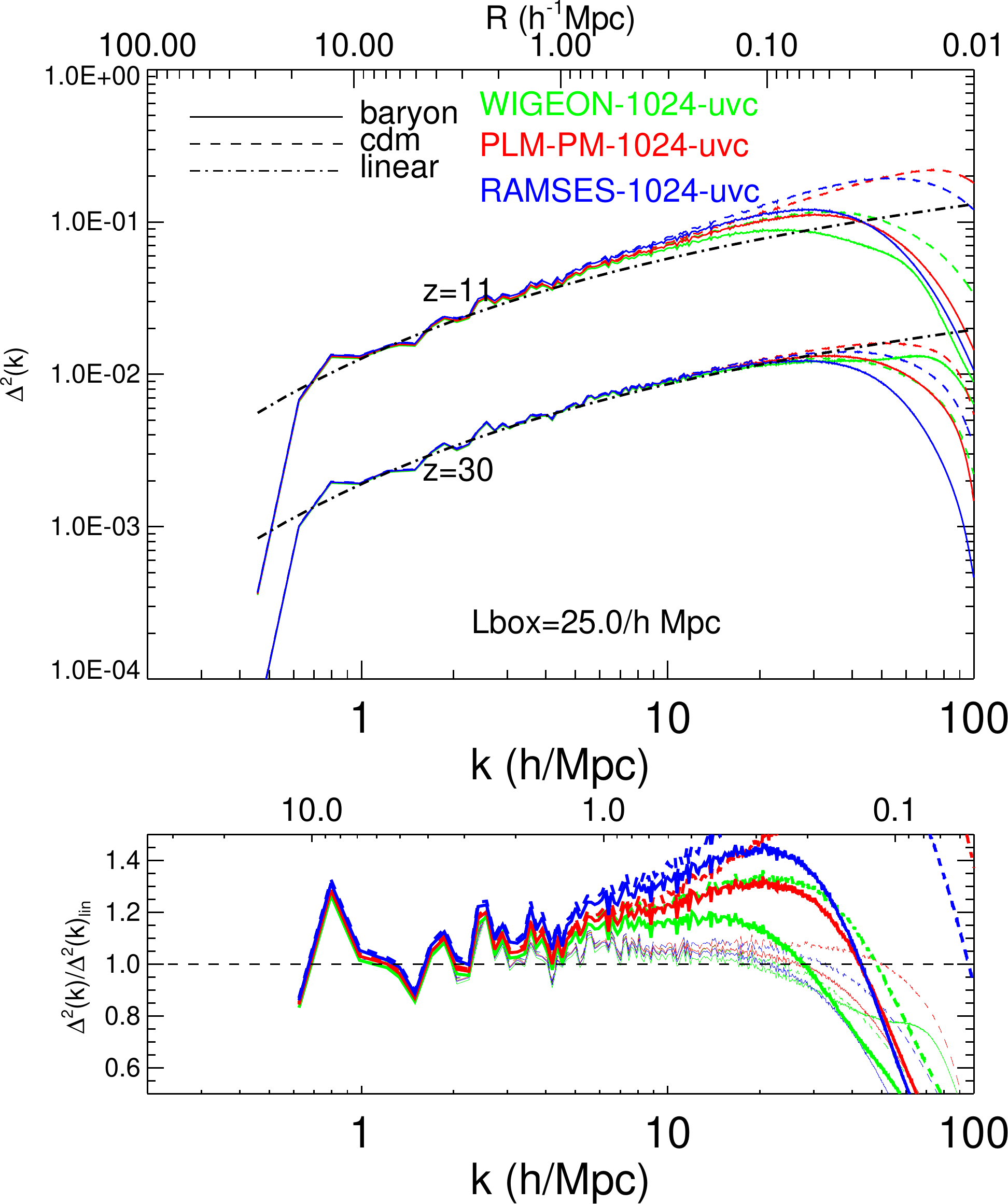}
\vspace{1.0cm}
\caption{Density power spectrum in our radiative simulations at redshift z = 20, 15, 11. Solid and dotted lines represent baryon and dark matter respectively. Dot dashed line represents dark matter power spectrum described by the linear evolution prediction. The bottom panel shows residuals from linear evolution prediction, thin and thick lines indicate results at $z=30$ and $z=11$ respectively.}
\label{fig:den_pow}
\end{center}
\end{figure}
To check whether the discrepancies on gas temperature difference between these three codes are caused by their dark matter distributions, we show dark matter and gas density power spectrum in Fig.~\ref{fig:den_pow}. We can see that on scale larger than 1 Mpc there is barely any difference between three sets of simulations for both baryonic and dark matter density power spectrum.  The power spectrum of dark matter in simulations deviates from the linear prediction slightly at $z=30$, with percent level difference at 200kpc. However, the power spectrum in simulations can be higher than the linear prediction by $20\%$ at $\sim 1\rm{Mpc}$ at $z=11$.

\section{density-temperature distribution in $256^3$ simulations}
\begin{figure}
\begin{center}
\vspace{-0.3cm}
\includegraphics[trim = 0mm 0mm 0mm 0mm, width=0.7\columnwidth]{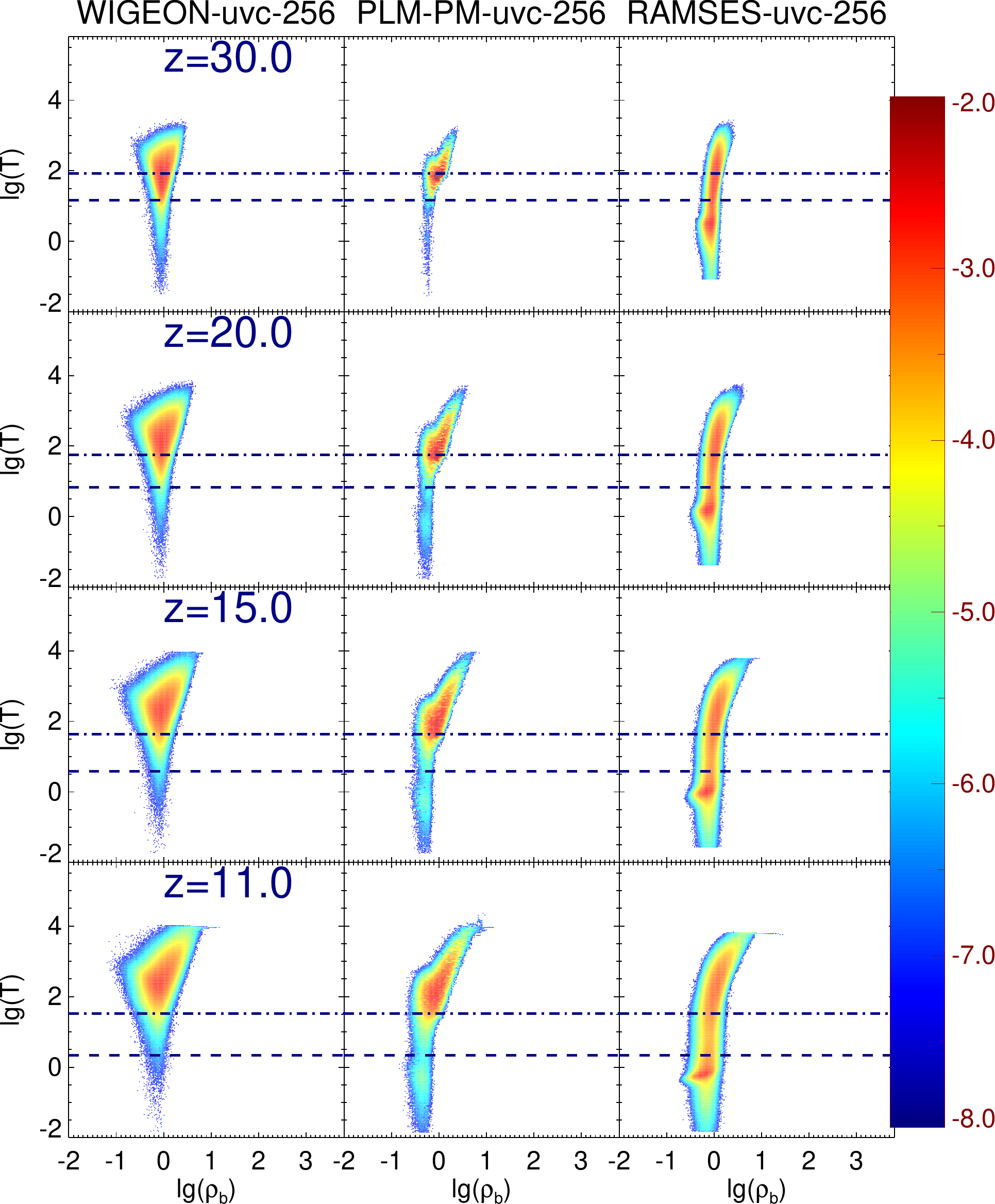}
\caption{Top: The volume-weighted distribution of the cosmic gas in the density-temperature plane at redshift $30, 20, 15, 11$ in our radiative simulations with resolution $256^3$ resolution; Left, middle and right column indicate simulations run with WIGEON, PLM-PM, and RAMSES, respectively; Dotted-dashed lines in each plot indicate the CMB temperature at corresponding redshifts; Dashed lines indicate the expected gas temperature due to expansion.}
\label{fig:tem_den_map_256}
\end{center}
\end{figure}
 To check the convergence of our work, we also plot the volume distribution of cosmic gas in the density-temperature phase plane in radiative simulations with $256^3$ resolution in Fig.~\ref{fig:tem_den_map_256}. Combined with the top panel of Fig.~\ref{fig:vol_dis_ada}, we can see that the evolution trend of thermal history is similar at different resolutions.


\label{lastpage}
\end{document}